\documentclass{aa}

\usepackage{natbib}
\usepackage{graphicx}
\usepackage{amssymb}
\usepackage{txfonts}
\begin{document}

  \title{Measuring surface magnetic fields of red supergiant stars
     \thanks{Based on observations obtained at the Télescope Bernard Lyot
(TBL) at the Observatoire du Pic du Midi, operated by the
Observatoire Midi-Pyrénées, Université de Toulouse (Paul Sabatier),
Centre National de la Recherche Scientifique of France.
}}

   \author{B. Tessore%
          \inst{1}, A. L\`ebre\inst{1}, J. Morin\inst{1}, P. Mathias\inst{2,3}, 
          E. Josselin\inst{1,2,3}, \and M. Aurière\inst{2,3}}

   \institute{LUPM, Universit\'e de Montpellier, CNRS, Place Eug\`ene Bataillon, 34095, France\\
              \email{benjamin.tessore@umontpellier.fr} \and 
     Universit\'e de Toulouse, UPS-OMP, Institut de recherche en Astrophysique et Plan\'etologie, Toulouse, France
\and
CNRS, UMR5277,  Institut de recherche en Astrophysique et Plan\'etologie, 14 Avenue Edouard Belin, 31400 Toulouse, France
              }

   \date{Received March, 2017; accepted April 19, 2017}

  \abstract
   {Red supergiant (RSG) stars are very massive cool evolved stars. Recently, a weak magnetic field was measured at the surface of $\alpha$~Ori and this is so far the only M-type supergiant for which a direct detection of a surface magnetic field has been reported.}
   {By extending the search for surface magnetic field in a sample of late-type supergiants, we want to determine whether the surface magnetic field detected on $\alpha$~Ori is a common feature among the M-type supergiants.}
   {With the spectropolarimeter Narval at Télescope Bernard-Lyot we undertook a search for surface magnetic fields in a sample of cool supergiant stars, and we analysed circular polarisation spectra using the least-squares deconvolution technique.}
   {We detect weak Zeeman signatures of stellar origin in the targets ${\rm CE~Tau}$, ${\rm \alpha^1~Her}$ and ${\rm \mu~Cep}$. For the latter star, we also show that cross-talk from the strong linear polarisation signals detected on this star must be taken into account. For ${\rm CE~Tau}$ and ${\rm \mu~Cep}$, the longitudinal component of the detected surface fields is at the Gauss-level, such as in $\alpha$~Ori. We measured a longitudinal field almost an order of magnitude stronger for ${\rm \alpha^1~Her}$. We also report variability of the longitudinal magnetic field of CE Tau and ${\rm \alpha^1~Her}$, with changes in good agreement with the typical atmospheric dynamics time-scales. We also report a non-detection of magnetic field at the surface of the yellow supergiant star ${\rm \rho~Cas}$.}
   {The two RSG stars of our sample, ${\rm CE~Tau}$ and ${\rm \mu~Cep}$, display magnetic fields very similar to that of $\alpha$~Ori. The non-detection of a magnetic field on the post-RSG star ${\rm \rho~Cas}$ suggests that the magnetic field disappears, or at least becomes undetectable with present methods, at later evolutionary stages. Our analysis of $\alpha^1$~Her supports the proposed reclassification of the star as an M-type asymptotic giant branch (AGB) star.}

                  \maketitle

\section{Introduction}\label{Intro}

With masses ranging from about $ 10~M_{\mathrm{\odot}}~\mathrm{to}~40~M_{\mathrm{\odot}}$, red supergiant (RSG) stars, meaning supergiants of M spectral-type, can be considered as the massive counterparts of asymptotic giant branch (AGB) stars.
These cool evolved stars are surrounded by a circumstellar envelope (CSE) that is rich in molecules and dust grains.

Similar to AGB stars, RSG stars are losing a large amount of mass, up to ${\rm 10^{-4}M_{\mathrm{\odot}}/yr}$. Asymptotic giant branch and red supergiant stars are therefore considered as important recycling agents of the interstellar medium.
To date, the physical processes contributing to their high mass-loss rates have not been fully identified, although in
turn, mass loss is an essential driver of stellar evolution, and thus a key ingredient to stellar evolution codes.

At photospheric level, both radiative hydrodynamics (RHD) simulations \citep{2009A&A...506.1351C} and interferometric observations \citep{2009A&A...508..923H} show that RSG stars have a small number of giant convective cells, in agreement with earlier theoretical predictions \citep{1975ApJ...195..137S}. Unlike pulsating AGB stars, RSG stars are irregular variables and their related luminosity variations can be assigned to the changes of the convective cells with time. 
Thus a mass loss triggered by pulsations, as occurring in for example Mira stars, is likely irrelevant for RSG stars.
Moreover, a significant amount of dust is only found far from the photosphere \citep{1994AJ....107.1469D} so that radiation pressure on dust grains cannot trigger a mass-loss event, as suggested for AGB stars \citep[see][]{2008A&A...491L...1H}. However, the surface temperature is low enough for molecules to form in the photospheric regions.
Regarding spectroscopic observations, \cite{2007A&A...469..671J} proposed that turbulent convective motions and radiation pressure on molecules may initiate a mass-loss event. They also recall that the magnetic field is often invoked in mass-loss mechanisms.

Magnetic fields have already been detected at several radii in the CSE of cool AGB stars and in the RSG star VX Sgr using ${\rm SiO,~H_2 O~and~OH}$ masers polarisation \citep[see for instance][and references therein]{2014IAUS..302..389V}.
Still, little is known about the surface magnetism of cool evolved stars, especially in RSG stars, although the new generation of spectropolarimeters have recently provided new information about the surface magnetism of supergiants of spectral type A-F-G-K.
\cite{2010MNRAS.408.2290G} report a detection rate of a surface magnetic field of one third in a sample of thirty yellow  supergiants of A-F-G-K type. However, they fail to detect a magnetic field at the surface of any of the three RSG stars, of spectral type M, in their sample, ${\rm \alpha~Ori,~\sigma~CMa~and~\alpha~Sco}$.
In a parallel study, \cite{2010A&A...516L...2A} detect a weak magnetic field, at the Gauss-level, at the surface of ${\rm \alpha~Ori}$ and this is so far the only RSG star for which a direct detection has been obtained.
A follow-up study also shows that this
surface magnetic field appears to vary on a monthly time-scale \citep{2013EAS....60..161B, 2013LNP...857..231P} in agreement with
the convective patterns time-scale \citep{2002AN....323..213F,2015EAS....71..243M}.

We collected and analysed high-quality spectropolarimetric data for a sample of three RSG stars and one yellow supergiant (hereafter, YSG).
In Sect. 2 we present the targets and the strategy of observations and in Sect. 3 we present our data analysis. In Sect. 4 we discuss our main results for individual objects and conclusions are presented in Sect. 5.

\section{Targets and observations}\label{TargetsObservations}
In this work we analyse circularly polarised spectra of three RSG stars, ${\rm \mu~Cep}$, ${\rm \alpha^1~Her}$, ${\rm CE~Tau}$, and of one YSG star, ${\rm \rho~Cas}$. 
We have chosen these three RSG stars for their similarities with the well known ${\rm \alpha~Ori}$, Betelgeuse; considered as our reference star throughout this paper. We also considered the YSG star ${\rm \rho~Cas}$, known to be in the post-RSG phase, to give a first glance at the magnetic properties of stars right after the RSG phase.

These targets belong to an observing program, initiated in 2015 at Télescope Bernard-Lyot (TBL, Pic du Midi France) with the Narval instrument, the twin of the ESPaDOnS spectropolarimeter  \citep{2006ASPC..358..362D}.
This large program is dedicated to the investigation of the surface magnetism of cool evolved stars located in the upper right part of the Hertzsprung-Russel diagram.
It includes objects such as M-type AGB stars, Mira stars, RV Tauri stars and our RSG targets plus ${\rm \alpha~Ori}$.
Table \ref{tab1:table 1} introduces the observed targets and presents their most relevant physical parameters for this study.
\begin{table*}[h!]
\centering
\begin{tabular}{llccccc}
    \hline
  \hline
  Target & Spectral type & $T_{\mathrm{eff}}$ (K) & $\log~g$ & $m_{\rm V}$ & ${M~(M_{\mathrm{\odot}})}$\\

  \hline
  ${\rm \mu~Cep }$ & M2eIa & 3750& -0.36 & 4.10&25\\
  ${\rm \alpha^1~Her }$ & M5Ib-II(+G8III+A9IV-V) &3280 & 0.0 & 3.35&2.5\\
  ${\rm CE~Tau }$&M2Iab-Ib  & 3510 & 0.0 & 4.30&15\\
  ${\rm \rho~Cas }$&G2Ia0e  & 6000& 0.25  & 4.59 &-\\

  ${\rm \alpha~Ori }$& M1-M2Ia-ab & 3780 & 0.08 & 0.42&15\\
    \hline
  \hline
\end{tabular}
\caption{Stellar parameters for each target. The visual magnitudes ($m_V $) are taken from the SIMBAD database.
Stellar parameters ($T_{\mathrm{eff}}$; ${log~g}$; ${M~(M_{\mathrm{\odot}})}$) are taken from \cite{2005ApJ...628..973L} for ${\rm \alpha^1~Her}$, CE Tau, ${\rm \mu~Cep}$ and ${\rm \alpha~Ori}$ (${\rm \alpha~Ori}$ is shown here for comparison) and from \cite{1983IzKry..66..130B} for ${\rm \rho~Cas}$. }

\label{tab1:table 1}
\end{table*}

We used Narval in circular polarisation mode to detect Zeeman signatures tracing the presence of a surface magnetic field. 
In polarimetric mode, Narval simultaneously acquires two orthogonally polarised spectra covering the spectral range from 375 nm to 1050 nm in a single exposure at a resolution of about 65\,000. 
A classical circular polarisation observation, hereafter a Stokes $V$ sequence, is composed of four sub-exposures between which the half-wave retarders (Fresnel rhombs) are rotated to cancel first-order spurious signatures.
The optimal extraction of spectra is performed with Libre-ESpRIT \citep{1997MNRAS.291..658D}, an automatic and dedicated reduction package installed at TBL and it includes wavelength calibration, continuum normalisation, and correction to the heliocentric rest frame.
The reduced spectra contain the normalised intensity (${I/I_{\rm c}}$, hereafter Stokes~$I$, with ${I_{\rm c}}$ the unpolarised continuum intensity), the normalised circular polarisation (${ V/I_{\rm c}}$, hereafter Stokes $V$) and their corresponding standard deviations as functions of wavelength. In the output spectra a null (${ N/I_{\rm c}}$) diagnosis that does not contain any physical informations about the target star is also included. However, in absence of spurious polarisation, the null diagnosis is featureless and statistically consistent with the noise, therefore indicating the good quality of polarimetric observations (see Sect. \ref{null}).

To detect Gauss-level magnetic field at the surface of supergiants it is necessary to reach a high signal-to-noise ratio (SNR) and therefore we need very long exposure time. However, in order to avoid saturation of the CCD, in the reddest orders, we have adopted the same strategy of observation as previous studies dedicated to cool evolved stars, which consists in co-adding many polarimetric Stokes $V$ sequences to ensure that the requested SNR, > 1000 per 2.6 ${\rm km.s^{-1}}$ velocity bin, is achieved. Considering the successful detection of a surface magnetic field on ${\rm \alpha~Ori}$ by \cite{2010A&A...516L...2A} and subsequent papers we took into account the following key points to reach a minimal detection threshold in Stokes $V$ observations better than the Gauss-level: (i) we have observed each target at several epochs separated by at least one month, to account for intrinsic variability; and (ii) we have collected between five and twenty five contiguous series of Stokes $V$ sequences for each target so that it is possible to average them to increase the SNR.
Table \ref{tab2:table 2} summarises the observations for each target.

 \begin{figure*}[h!!!!]
         \begin{minipage}{\columnwidth}
	\includegraphics[width=\columnwidth]{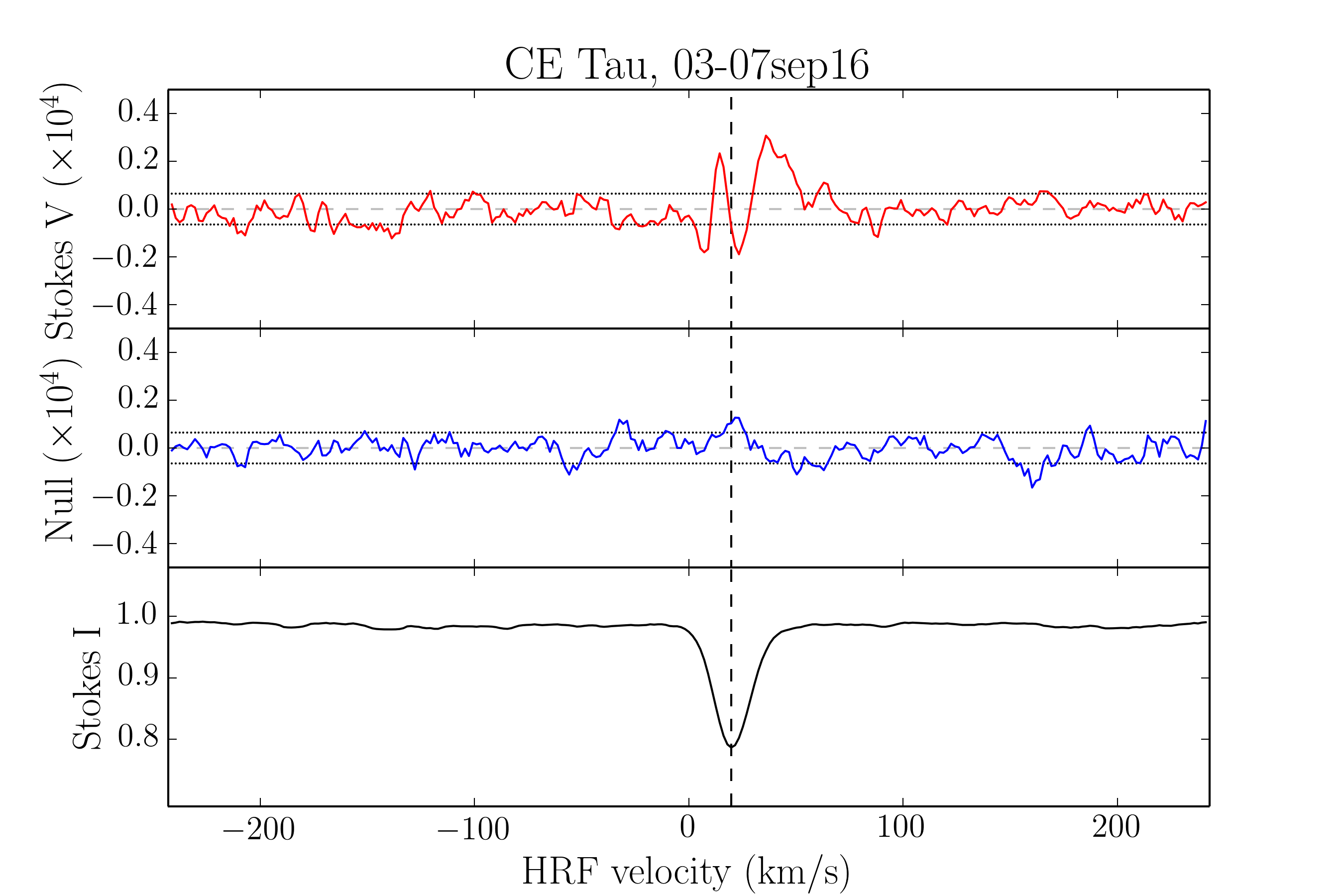}

	\includegraphics[width=\columnwidth]{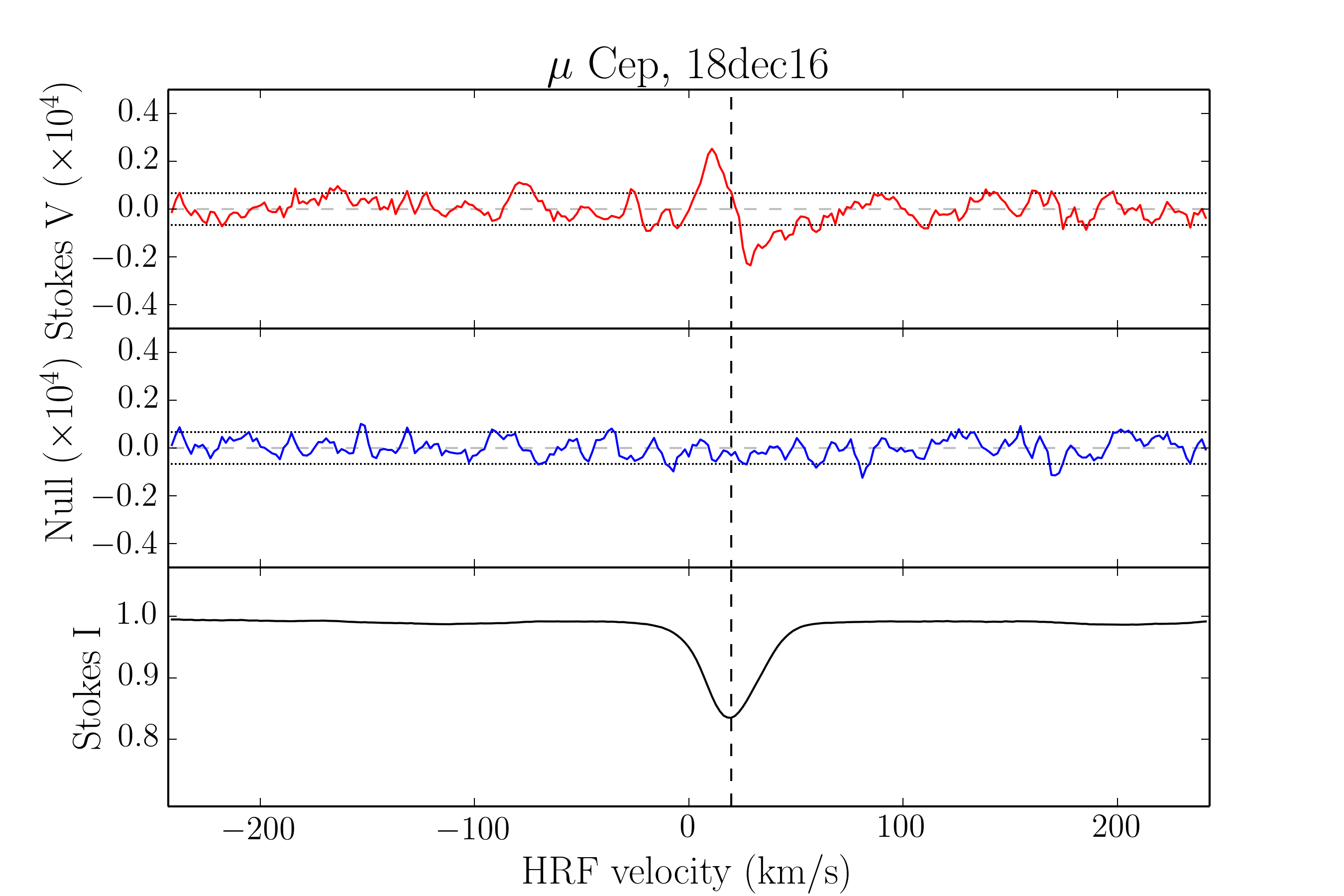}

        \end{minipage}
        \begin{minipage}{\columnwidth}
	\includegraphics[width=\columnwidth]{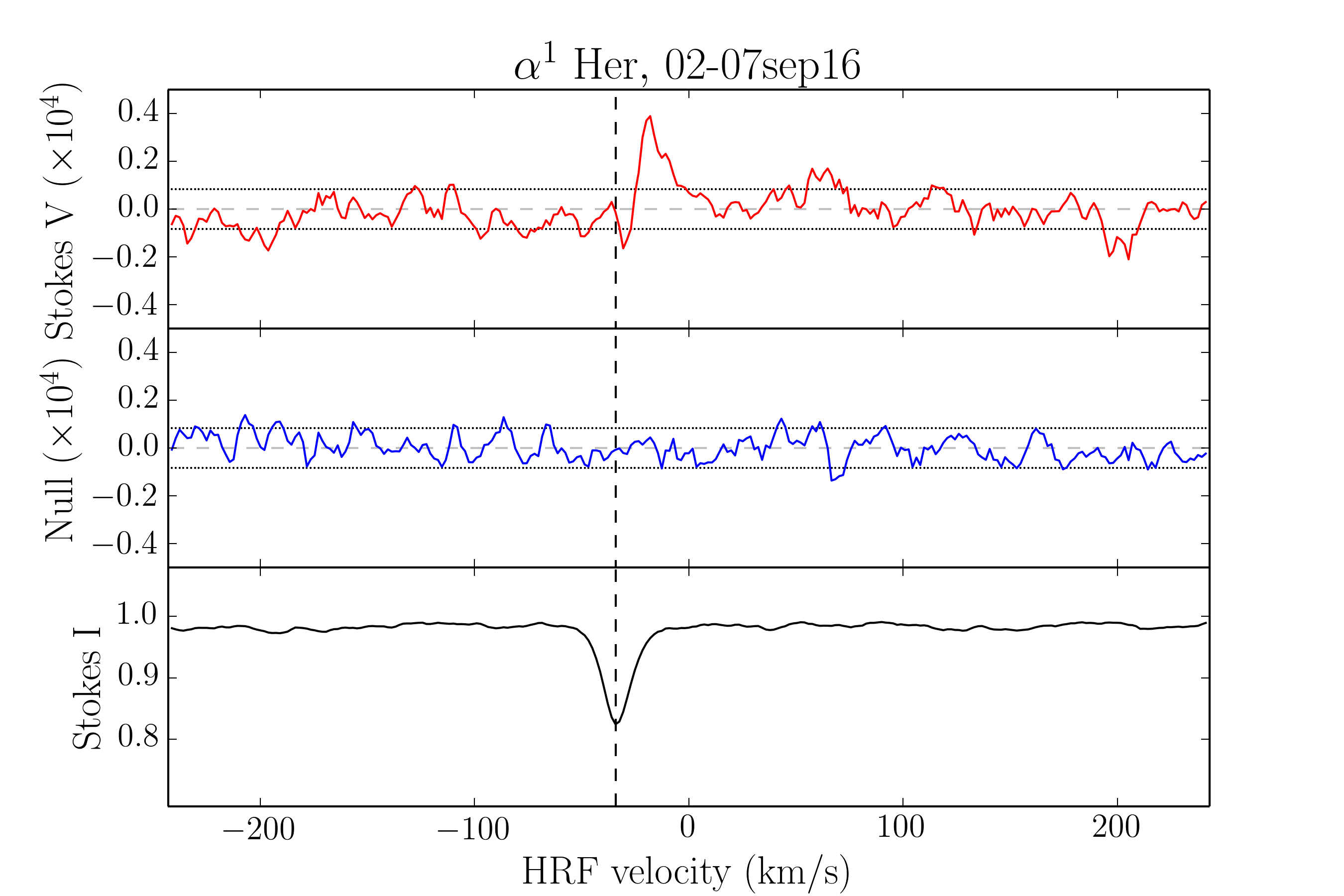}

	\includegraphics[width=\columnwidth]{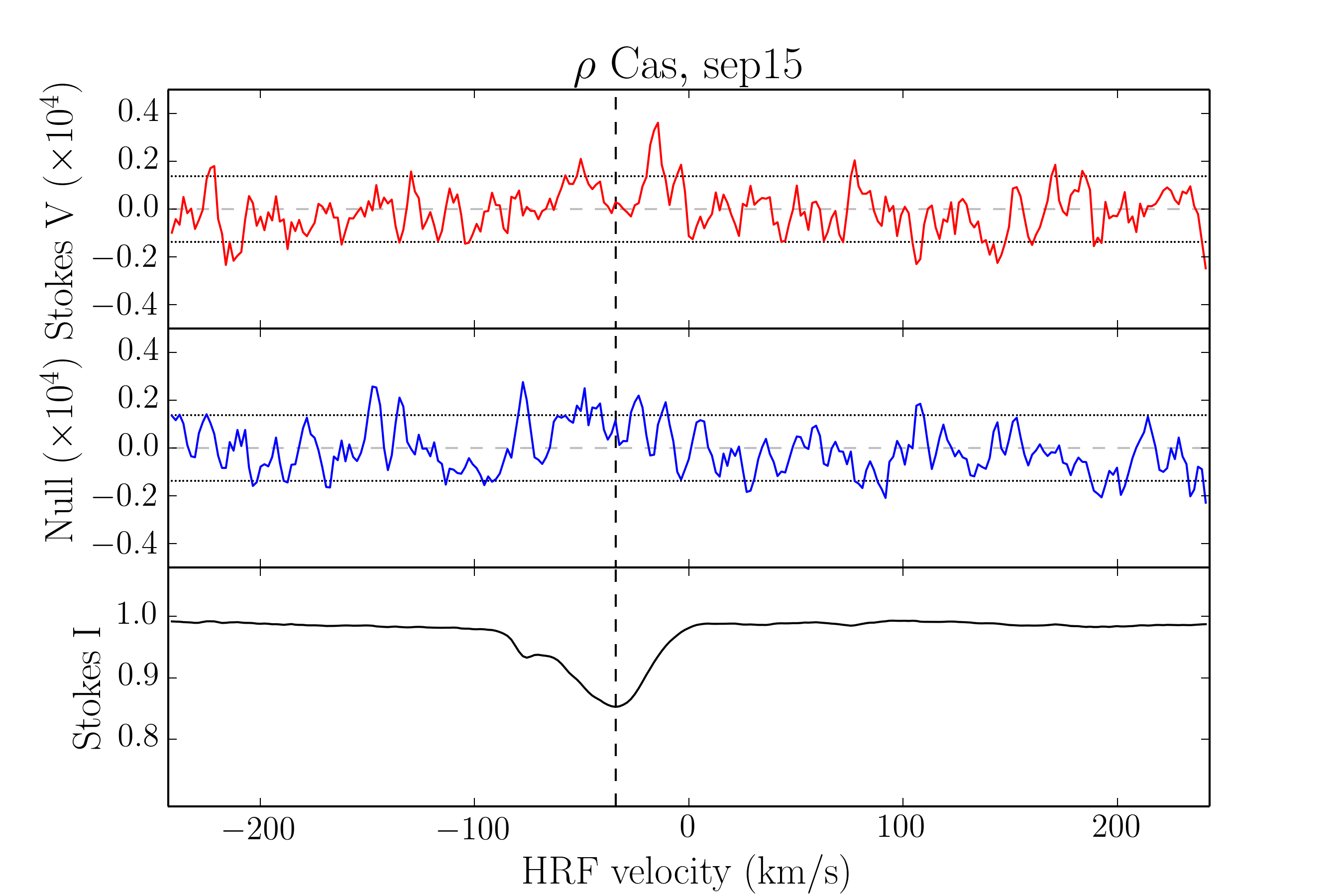}	
	 \end{minipage}

    \caption{Highest SNR LSD profiles. For each panel: \textit{\textbf{Top}.} Stokes $V$ in red, \textit{\textbf{ Middle}.} null diagnosis in blue and \textit{\textbf{ Bottom}.} Stokes $I$ in black. The doted horizontal lines represent the $\pm~1\sigma$ level and the dashed horizontal lines the 0 position. The vertical line marks the radial velocity of the star in the heliocentric rest frame. The profile of $\mu$~Cep is free from linear polarisation to circular polarisation cross-talk (see Sect. \ref{null} and Sect. \ref{CrossTalk})}

     \label{fig:figure 1}
 \end{figure*}

\begin{table*}
\centering
\begin{tabular}{lcccccc}
  \hline
    \hline
  Target &  Julian dates (HJD-2,400,000.5) & Dates (yyyy-mm-dd) & Exposure time (s)& Sequences & SNR peak (/2.6 ${\rm km.s^{-1}}$ bin)\\

  \hline
  ${\rm \mu~Cep }$  & 57\,213.1 & 2015-07-09 &200 & 16 & 1740\\
        & 57\,266.9 & 2015-09-01 & 200& 25 &1550\\
        & 57\,309.9 & 2015-10-14 & 200& 25 & 1391\\
        &  57\,337.9 & 2015-11-11 & 200 & 25 &1542\\
        &  57\,740.8 & 2016-12-18$^{a}$ & 200& 8 &1695\\
       &  57\,760.8 & 2017-01-07 & 200 & 8 &1749\\

  \hline
    \hline
  ${\rm \alpha^1~Her }$ & 57\,093.1 & 2015-03-11 & 120 & 16 &2122\\ 
        & 57\,214.9 & 2015-07-11 & 120  & 16 & 2069\\
         & 57\,271.8 & 2015-09-06 & 120  & 9&1970\\

                  & 57\,633.9 & 2016-09-02 & 120  & 5&1842 \\

                  & 57\,634.8 & 2016-09-03 & 120  & 11& 1632\\

                  & 57\,636.9 & 2016-09-05 & 120  & 5& 1177\\

                  & 57\,637.9 & 2016-09-06 & 120  & 7& 1537\\

                  & 57\,637.9 & 2016-09-07 & 120  & 7& 1302\\

   \hline
     \hline
  ${\rm CE~Tau }$  & 57\,087.9 & 2015-03-06 & 300 & 16 & 1004\\
  & 57\,635.2 & 2016-09-03 & 300 & 16& 1265\\
  & 57\,639.2 & 2016-09-07 & 300 & 10& 1461\\
  & 57\,676.2 & 2016-10-14 & 300 & 16& 1501\\
  & 57\,741.0 & 2016-12-18 & 300 & 16& 1501\\

  \hline
    \hline
  ${\rm \rho~Cas }$       & 57\,246.1 & 2015-08-11 &  400 & 14 & 956\\
          & 57\,263.1 & 2015-08-28 &  400& 16 & 1008\\
  & 57\,273.1 & 2015-09-07 & 400& 21 & 1033\\
                       & 57\,274.1 & 2015-09-08 & 400& 19 &1069 \\
                         & 57\,276.1 & 2015-09-10 & 400& 5 & 752\\

  \hline
 \hline
\end{tabular}
\caption{Journal of Narval observations. The first column recaps the name of the observed targets and the second and the third columns give the dates of the observations (Heliocentric Julian date and Gregorian calendar, respectively). The fourth column displays the total exposure time for a single Stokes $V$ sequence (combination of four sub-exposures). The fifth column gives the number of Stokes $V$ sequences collected for each observational date. Finally the last column gives the mean maximum SNR reached in an individual Stokes $V$ spectrum. The exponent $^{a}$ for the star $\mu$~Cep, indicates the technical observations presented in Sect. \ref{CrossTalk}.}
\label{tab2:table 2}
\end{table*}

\section{Data analysis}\label{DataAnalysis}

\subsection{Methods}\label{Methods}
 
Because the Zeeman signatures we want to detect in Stokes $V$ sequences are very faint, with amplitudes relative to the continuum typically below $10^{-4}$, we need to reach very high SNR and this cannot be achieved in one spectral line. Hence, to get a good spectral lines diagnosis, so as to detect those faint amplitude signals, we have used the least-squares deconvolution (LSD) method \citep{1997MNRAS.291..658D}.
It assumes that the observed spectrum is the convolution of a weighted Dirac comb, parametrised by the Landé factor, the depth, and the central wavelength defining the spectral lines, with a mean line profile. Then the LSD algorithm solves the inverse problem, finding the mean line profile knowing the observed spectrum.
Therefore the LSD method is a multi-line technique that extracts a mean line profile, also called an \textit{LSD profile}, from thousands of spectral lines, the typical value for cool stars, referenced in a mask file (see below).
Hence the SNR is drastically amplified roughly proportionally to the square root of the number of lines used. The multiplex gain, defined as the ratio between the SNR in the LSD profile and the spectrum SNR peak is about twenty.

The LSD profile is computed by means of a digital mask gathering the intrinsic parameters of thousands of atomic lines provided by the Vienna Atomic Line Database \citep[VALD;][]{1999A&AS..138..119K}. 
This mask is computed  with a Kurucz model atmosphere with solar abundances \citep{2005MSAIS...8...14K}. 
For each target, this mask depends on the assumed stellar parameters, $T_{\mathrm{eff}}$ and $\log g$, and more specifically on the details of the underlying model atmosphere such as chemical composition and opacities. 
For our three RSG stars, which have similar stellar parameters (see Table~\ref{tab1:table 1}), we used the same mask as in \cite{2010A&A...516L...2A} for ${\rm \alpha~Ori}$ : $T_{\mathrm{eff}} = 3750~\mathrm{K}, \log g = 0.5 $. 
For our YSG star ${\rm \rho~Cas}$, we have selected a model atmosphere with stellar parameters: $T_{\mathrm{eff}}=6000~\mathrm{K}$ and $\log g=0.5$.\\

An LSD profile is normalised by three free parameters: the equivalent depth ($\overline{d}$), the equivalent Landé factor ($\overline{g}$) and the equivalent wavelength ($\overline{\lambda}$).
The LSD method relies on two main assumptions: (i) all the lines have the same shape, weighted by intrinsic parameters, namely central depth relative to the unpolarised continuum, Land\'e factor and central wavelength; and (ii) the lines add up linearly.
We also chose to avoid lines with circumstellar and/or chromospheric contributions, meaning H, He, and resonance lines, so as to respect assumption (i).
To increase the SNR, again, for each star we averaged the LSD profiles of contiguous series of Stokes $V$, when the dates of observations were close enough to result in similar Stokes $I$ profiles. 

However, the spectra of very cool stars contain a large number of molecular lines which are not yet included in our LSD analysis, such as TiO bands. 
Therefore the spectral lines of RSG stars are strongly blended and these atomic and molecular blends can be a real problem for performing LSD analysis because assumption (ii) may be broken. 
However, several analysis on cool stars show that LSD gives accurate results even with strong molecular blend typical of early- and mid-M spectral types \citep{2008MNRAS.384...77M,2010A&A...516L...2A,2014A&A...561A..85L}.                                                                                               
To avoid the over-representation of weak lines in the mask and to ensure a robust analysis with LSD, only lines with depth over 40\% of the continuum were kept \citep{1997MNRAS.291..658D}, resulting in an atomic mask with ${\rm 12000}$ lines over the observed spectral range.

We adopted the same statistical criteria as in \cite{1997MNRAS.291..658D}, widely used in the stellar spectropolarimetry community, that result in a "definite detection" (DD), "marginal detection" (MD) and "no detection" (ND) if the detection probability is, respectively, over 99.99\%, over 90.0\%, and lower than 90.0\%. We also applied these criteria to the null profiles to assess whether they display any significant features. In case for which the statistics lead to a "no detection" in the null profiles we attribute a Zeeman origin for the Stokes $V$ signals, based on the successful detection of very weak circularly polarised signatures in the RSG star $\alpha$~Ori \citep{2010A&A...516L...2A} and in cool giants \citep[see for instance ][]{2013BlgAJ..19...14K,2014arXiv1410.6224S,2014A&A...561A..85L,2015A&A...574A..90A}. However, in the case of a "definite detection" of a signal in the null profiles we performed an in-depth study of this signal.

\subsection{Detections}\label{Detections}
Figure \ref{fig:figure 1} shows the LSD profiles with the highest SNR among all the observations of each target.
We applied the LSD statistical tools to these averaged signals.\\
For $\rho$ Cas we obtain a "no detection" both in Stokes $V$ and in the null. For CE Tau and $\alpha^1$ Her we obtain a "definite detection" in Stokes~$V$ and a "no detection" in the null. \\More problematic was the case of $\mu$ Cep, where we obtain a "definite detection" both in Stokes $V$ and in the null in most of our observations.
\\The complete statistics for each target and each LSD profile are presented in Table \ref{tab3:table 3}.

\begin{table*}
\centering
\begin{tabular}{lccccccc}
  \hline
    \hline

  Target & Obs. date & $\overline{\lambda}$ & $\overline{g}$ & $\overline{d}$ &$B_{\ell}\pm~\sigma ~(G)$ & Detection (Stokes $V$) & Detection ($null$) \\
  \hline
  ${\rm \mu~Cep }$  & 2015-07-09* & 772 & 1.14 & 0.66 & - & - & DD \\ 
                   & 2015-09-01* & 772 & 1.14 & 0.69 &- & - & DD\\ 
                   & 2015-10-14* & 772 & 1.14 & 0.69 &- & -& DD\\ 
                   & 2015-11-11* & 772 & 1.12 & 0.69 &- & -& DD\\
                   & 2016-12-18 & 772 & 1.24 & 0.69 & $1.0 \pm 0.3$ $\dagger$ & DD$\dagger$ & ND\\
                   & 2017-01-07 & 772 & 1.16 & 0.69 & $ 1.3 \pm 0.3$ $\dagger$ & DD$\dagger$ & ND\\

                   \hline
                     \hline

  ${\rm \alpha~Her }$ & 2015-03-11 & 772 & 1.19 & 0.68 &$-5.8 \pm 0.4$ & DD & ND\\
                     & 2015-07-11 & 772 & 1.19 &  0.66 &$ -7.4 \pm 0.3$ &  DD & ND\\
                   & 2015-09-06 & 772 & 1.20 & 0.66 &$ -7.6 \pm 0.5$ & DD & ND\\
                  & 2016-09-02/03/05/06/07 & 772 & 1.27 & 0.67 &$ -2.7 \pm 0.3$ & DD & ND\\

                   \hline
                     \hline

  ${\rm CE~Tau }$& 2015-03-06 & 772 & 1.20 & 0.69 &$-1.2 \pm 0.3$ &  MD & ND\\
    & 2016-09-03/07 & 772 & 1.20 & 0.68 &$ -1.7 \pm 0.2$ &  DD & ND\\
  & 2016-10-14 & 772 & 1.13 & 0.69 &$ -1.2 \pm 0.2$ &  DD & ND\\
  & 2016-12-18 & 772 & 1.15 & 0.69 &$ -2.7 \pm 0.5$ &  DD & ND\\

                   \hline
                     \hline

  ${\rm \rho~Cas }$& 2015-08-11/28 & 570 & 1.30 & 0.64 & - & ND & ND\\
                  & 2015-09-07/08/10& 570 & 1.30 & 0.62 & - & ND & ND\\

  \hline

\hline
\end{tabular}
\caption{Longitudinal magnetic fields computed for each observation. The first and the second column give the targets and the observation dates, respectively. The third, fourth  and fifth column give the LSD parameters introduced in  Sect. \ref{Methods} The last columns give the values of the longitudinal magnetic field and the $1\sigma$ level for each LSD profile, and the associated statistical threshold flag (as described in Sect. \ref{Methods}). The * for $\mu$ Cep observations indicates that they are plagued by spurious polarisation (see Sect. \ref{CrossTalk}) and therefore we did not compute the longitudinal magnetic field. The $\dagger$ indicates that the LSD statistics and the estimation of $B_{\ell}$ were performed on the cleaned profiles of $\mu$ Cep (see Sect. \ref{CrossTalk} and \ref{Bl}).}
\label{tab3:table 3}
\end{table*}

\subsection{null profiles analysis}\label{null}
As explained in the Sect.~\ref{Methods}, in addition to the Stokes $V$ profile, the LSD method provides a null diagnosis profile.
Although indicative, the absence of signature in the null diagnosis cannot be taken as a definite proof of the absence of contamination of the Stokes~$V$ by spurious polarisation, as shown in \cite{2013A&A...559A.103B}. 
However, for a stabilised echelle spectropolarimeter without insertable optical element and with an optimised calibration procedure and a reduction software such as Narval, the effects mentioned by \cite{2013A&A...559A.103B} are largely mitigated.

The changing sky conditions or the variability of the star during the four sub-exposures, for instance, may induce a spurious signal and the associated null signal, that generally leads to rejection of the corresponding observation. 
On the other hand, though some instrumental or data-reducing effects might affect Stokes~$V$ without affecting the null signal as shown for FORS/VLT by \cite{2013A&A...559A.103B}, the absence of signature in the null diagnosis is considered as an indication of a sound measurement. Evidence of this kind of pollution of Stokes~$V$ results has never been shown for Narval nor ESPaDOnS up to now. Actually, Stokes~$V$ signals as weak as those observed for RSG stars have been currently observed for red giant stars and A-type stars, with insignificant associated null signals, including two stars whose magnetic fields are found to vary with a period that has been independently spectroscopically detected by other authors, Vega \citep[A0V, rotation period about 0.7 d,][]{2010A&A...523A..41P} and, Pollux \citep[K0III, period about 590 d,][]{2014IAUS..302..359A}. These observations clearly show that Stokes~$V$ data of Vega and Pollux are dominated by a stellar signal and that it is unlikely that the Stokes~$V$ of RSG stars are significantly polluted by an unknown instrumental problem when the associated null profiles are featureless. 
Moreover, the coexistence of detections and non-detections of weak 
magnetic fields among RGB and AGB stars of similar properties observed under equivalent 
conditions, and the fact that the detections gather in "magnetic strips" 
\citep{2015A&A...574A..90A,2017arXiv170310824C} further supports our ability to 
reliably detect weak magnetic fields with Narval using the null parameter as a 
quality check.

In the following paragraphs we further investigate the possible causes of the signals detected in the null profiles, focusing on the case of ${\rm \mu~Cep}$.
Firstly, we note that the majority of our $\mu$~Cep profiles are systematically plagued by a definite null signal, sharing a similar shape as the Stokes~$V$ signal, although our observations have been collected in diverse weather conditions, including photometric conditions, and at a range of airmasses. This rules out the explanation related to poor and/or changing sky conditions. Secondly, RSG stars are not known to undergo significant variations on a time-scale of a few hours on which our observations are generally collected. We also note that the timespan over which a sequence of spectra is acquired varies from a few hours to a few days whereas we always observe a clear signal in the null profile.

\citet{2016MNRAS.457..580F} found that for young cool stars, in case of very poor SNR, a null signal could occur due to the very noisy blue part of the spectrum. In spite of the high peak SNR ratio of our observations, due to the intrinsic spectral energy distribution of RSG stars, the bluest orders of our spectra do have a low SNR. Looking at observations of our Large Program of the RSG star ${\rm \alpha~Ori}$ we found that few of the LSD Stokes $V$ profiles are associated with a clear signal in the null, whereas for ${\rm \mu~Cep}$ the majority of the Stokes $V$ profiles are associated with a definite null signal except for the last observation of the star dated from 2017 January 7. We removed the blue part of the spectrum, up to $\lambda~=$ 500 nm, of ${\rm \mu~Cep}$ before performing the LSD analysis but this resulted in no noticeable improvement, thus ruling out Folsom's hypothesis.\\
In the next section we explore another possible source of spurious Stokes~$V$ polarisation that may lead to a signature in the null profile: the presence of strong linear polarisation of stellar origin.

\subsection{Incidence of linear polarisation on our Stokes~$V$ observations}\label{Lin2Circ}
\subsubsection{Cross-talk of ESPaDOnS and Narval}\label{CTnarv}
In their study of a sample of A-F-G-K-M supergiant stars, \cite{2010MNRAS.408.2290G} consider the effect of stellar linear polarisation on circular polarisation observations, they noted that prior 2009 ESPaDOnS instrument was known to suffer from cross-talk effect, mainly due to the Atmospheric Dispersion Corrector (ADC) and that this problem was almost solved when a new ADC was installed in 2010, reducing the cross-talk level in ESPaDOnS from 5\% to 0.6\% \citep{2010SPIE.7735E..4CB}. Besides, \cite{2010MNRAS.408.2290G} disregarded the possibility that strong linear polarisation signals exist in the spectra of supergiant stars, because this phenomenon was not known at that time yet, and that any such signal would be too low to give any significant contribution to the measured circular polarisation. Moreover, the repeatability of their measurements before and after the ADC changing made them confident with the stellar origin of the circular polarisation signals.
Narval, the twin of ~ESPaDOnS, suffers also from cross-talk between linear and circular polarisation which has to be 
taken into account when investigating very small polarisation levels.
The cross-talk on Narval has been measured directly on the sky  \citep{2012MNRAS.426.1003S} from 2009 observations and estimated to be at most of the order of 3\%.
A subsequent characterisation of the cross-talk performed in 2016 demonstrated that this level remains stable at about 3\% (Mathias et al. in preparation).
Recently, \cite{2016A&A...591A.119A} have discovered the complex linearly polarised spectrum of ${\rm \alpha~Ori}$. This spectrum, dominated by scattering processes, has a non-magnetic origin and is revealed by strong features detected in Stokes U and Stokes Q spectra in individual lines as well as in the LSD mean profiles. 
Within the framework of our observing program, we also obtain linearly polarised spectra of some of our targets. We detect linear polarisation in the spectra of CE Tau, about 0.01\%, and ${\rm \mu~Cep}$, about 0.1\%, that is much higher than circular polarisation and in contrast with the study of \cite{2010MNRAS.408.2290G} it cannot be neglected any more. The detailed analysis of the linearly polarised spectra of CE Tau and $\mu$ Cep will be discussed in a forthcoming paper.

\subsubsection{Investigating cross-talk from linear polarisation}\label{CrossTalk}

\begin{figure}%

        	\includegraphics[width=\columnwidth]{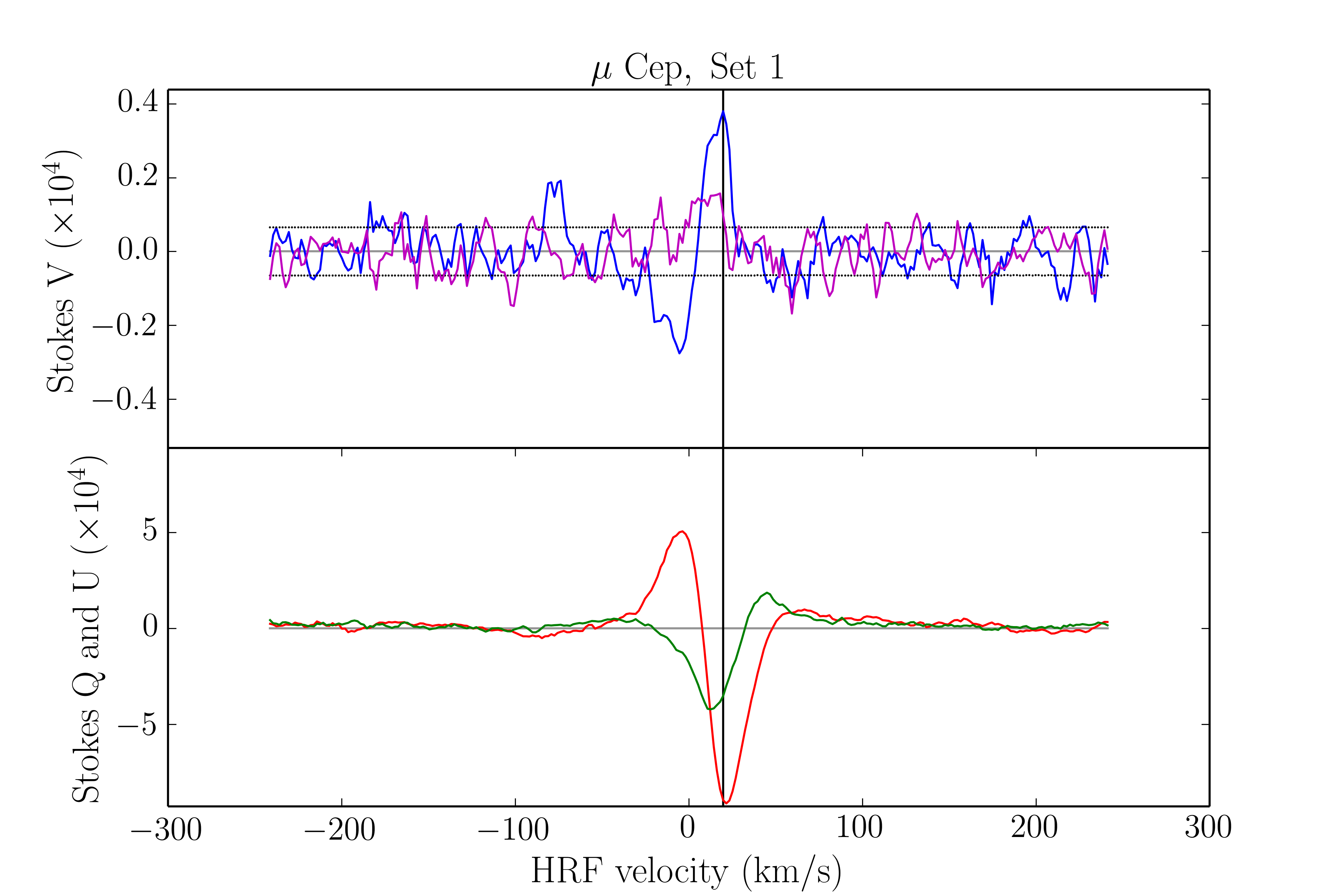}

	\includegraphics[width=\columnwidth]{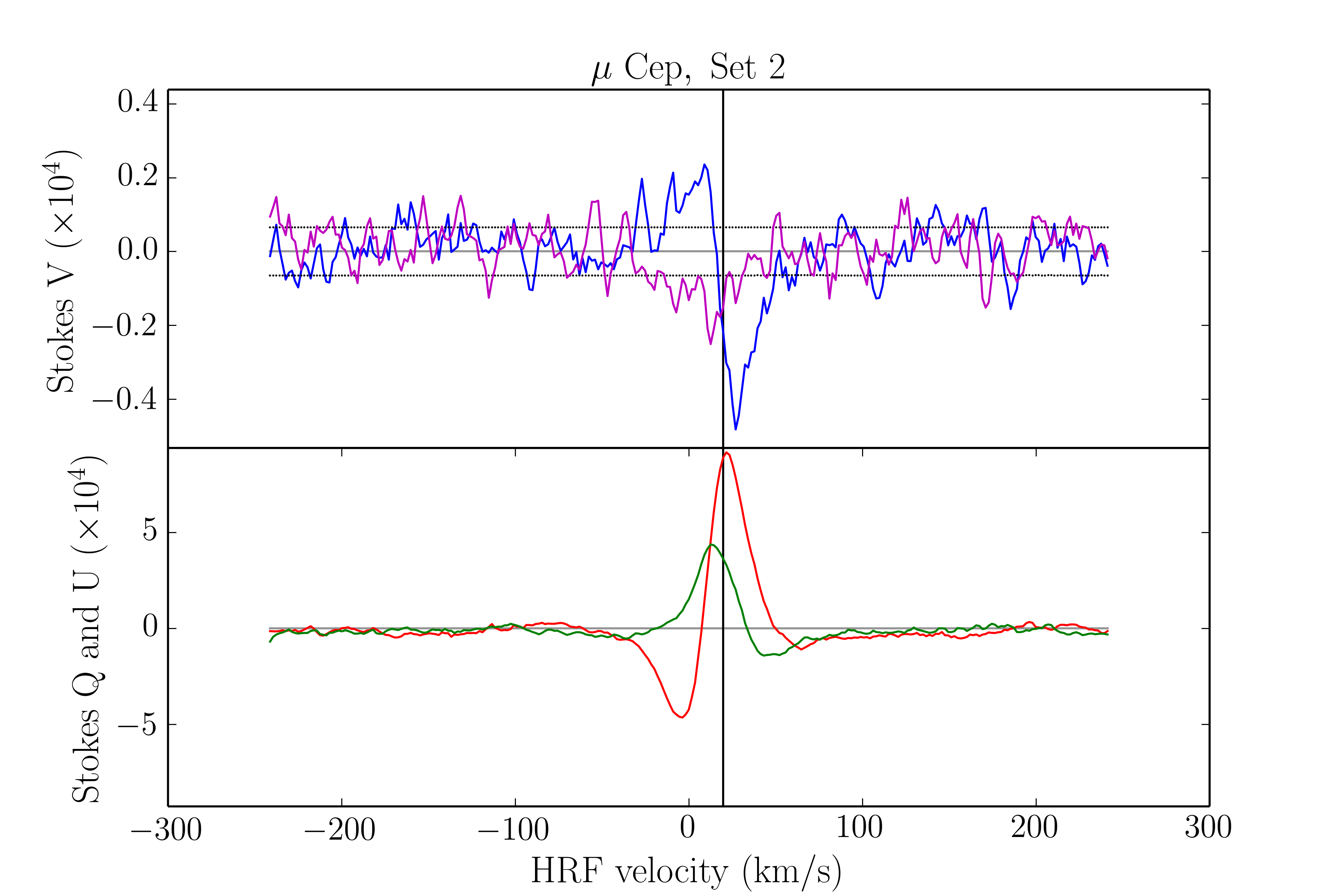}

    \caption{Mean Stokes $V$ profiles (blue), null diagnosis (purple), Stokes Q (red) and Stokes U (green) profiles for the two sets described in Sect. \ref{CrossTalk}. Upper panel shows set 1 and lower panel show set 2. The horizontal dotted lines represent the 1$\sigma$ level, the horizontal solid line the 0 level and the vertical line indicates the velocity of the star in the heliocentric rest frame.}

    \label{fig:figure 2}
  \end{figure}

We estimated the ratio between the level of circular polarisation to the total amount of linear polarisation for our targets and found roughly 0.1 for CE~Tau and ${\rm \alpha~Ori}$ and 0.01 for ${\rm \mu~Cep}$, suggesting that the detected Stokes~$V$ signature of $\mu$~Cep may be polluted by cross-talk from strong linear polarisation.
To clearly disentangle genuine circular polarisation signals of stellar origin from spurious signals due to cross-talk, we conducted a special observation run of $\mu$~Cep with the Narval instrument dated from 2016 December 18. We acquired two consecutive sets of observations, set 1 and set 2, both containing eight Stokes $V$ sequences and four Stokes U and four Stokes Q sequences. The polarimeter position angle was rotated by -90$^{\circ}$ for set 2 observations with respect to set 1, nominal polarimeter position angle, observations. Except for that, all observations were made in the same fashion as the observations described in Sect.~\ref{TargetsObservations}, meaning with the same exposure time and with ADC. \\Figure \ref{fig:figure 2} presents these two sets.
As a consequence of the -90$^{\circ}$ rotation of the instrument position angle, the orthogonal states of linear polarisation, Stokes~Q and Stokes~U, change signs from set 1 to set 2, and so do spurious Stokes~$V$ signals related to cross-talk. On the contrary, genuine circular polarisation if present in Stokes~$V$, remains unchanged.

Following \cite{2009PASP..121..993B} we then computed the half-sum of the Stokes~$V$ profiles of each set which results in a cross-talk-free Stokes~$V$ profile. On the contrary, the half-difference of the Stokes~$V$ profiles of each set cancels out true circular polarisation signals while preserving only the spurious signal caused by linear-to-circular polarisation cross-talk, allowing us to clearly distinguish the two components of the observed Stokes~$V$ profile. The situation is opposite for linear polarisation, the final Stokes~Q and Stokes~U profiles are obtained with the half-differences of the Q and U profiles measured in each set.

\begin{figure}%

	\includegraphics[width=\columnwidth]{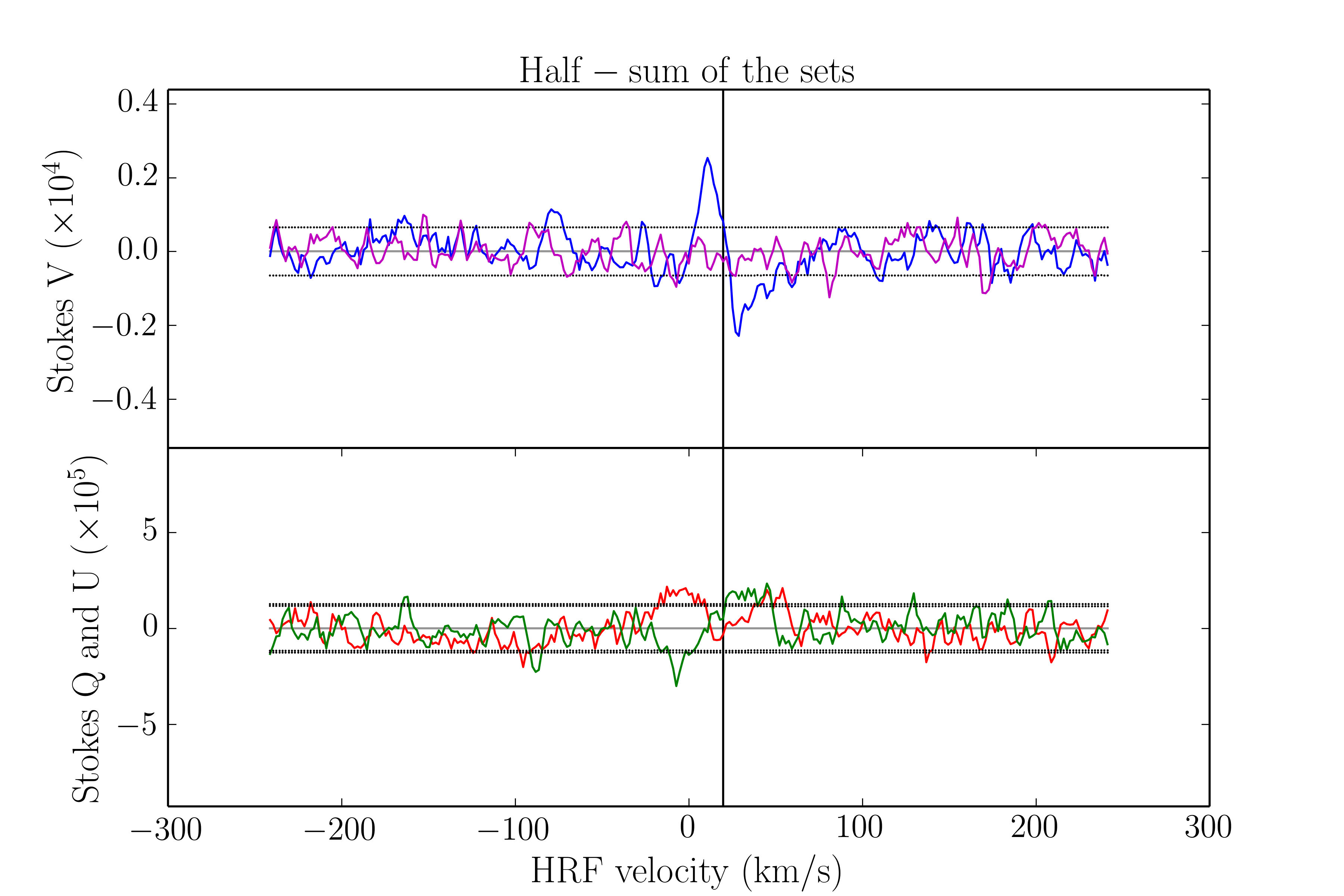}

	\includegraphics[width=\columnwidth]{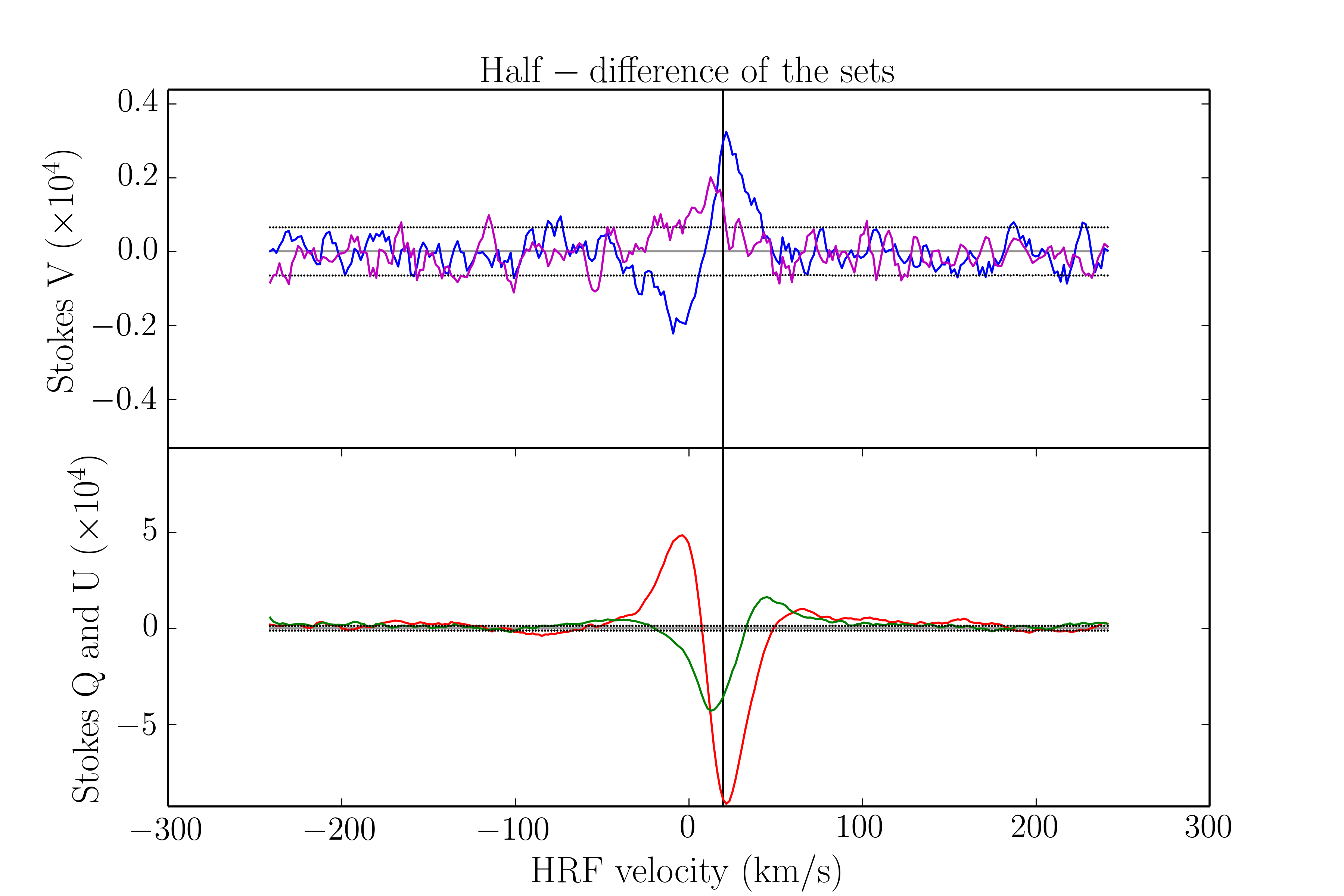}

    \caption{Mean Stokes $V$ profiles (blue), null diagnosis (purple), Stokes Q (red) and Stokes U (green) profiles obtained after adding/subtracting the two sets described in Sect. \ref{CrossTalk}. Upper panel shows the half-sum mean and the lower panel the half-difference mean. The horizontal dotted lines represent the 1$\sigma$ level and the horizontal solid lines the 0 level.}

    \label{fig:figure 3}
  \end{figure}
Figure \ref{fig:figure 3} shows the LSD profiles resulting from the half-sum and the half-difference between each set. Firstly, we clearly see that, as expected, the half-sum (Fig. \ref{fig:figure 3} upper panel) cancels the Stokes Q and U linear polarisation signals perfectly down to the noise level, demonstrating that the line profile is stable over a few hours and that it is safe to average several polarimetric sequences. The resulting cleaned Stokes $V$ signal is unambiguously above the noise level, "definite detection (DD)" according to our statistical criterion, whereas the null diagnosis is reduced down to the noise level, corresponding to "no detection  (ND)". Secondly, we see that the half-difference (Fig. \ref{fig:figure 3} lower panel) results in mean Stokes~Q and Stokes~U profiles whereas the mean Stokes~$V$ signal is well beyond the noise level, corresponding to "definite detection". The null diagnosis in this case is slightly beyond the noise level but below the statistical detection threshold, corresponding to "no detection". 

We then disentangled two components of Stokes~$V$. One that is clearly due to cross-talk and corresponds to about 3\% of the related linear polarisation (lower panel of Fig. \ref{fig:figure 3}); it is consistent with the known level of cross-talk previously measured for the Narval instrument on standard polarimetric targets.  The other component has about the same intensity, when the related linear polarisation is at the noise level (upper panel of Fig. \ref{fig:figure 3}): it is about 70 times more than what could be expected for a cross-talk effect from the related linear polarisation, and then cannot be related to this effect.  With this approach we therefore manage to detect unambiguously a Stokes V signal of stellar origin in the LSD profile of $\mu$ Cep.

Moreover, following \cite{2009PASP..121..993B}, the pure cross-talk Stokes~$V$ profile resulting from the half-difference can be approximated by\textbf{$^{\footnotemark[1]}$}:
\begin{equation}
 V_{\mathrm{cross-talk}}= \alpha Q + \beta U, 
\end{equation}

\footnotetext[1]{As explained in CFHT website, the reduction package Libre-ESpRIT inverts the sign of Stokes~U. For that reason, in this paper, we use -U in place of U. See \url{http://www.cfht.hawaii.edu/Instruments/Spectroscopy/Espadons/} for further information.}

with Q and U the orthogonal states of linear polarisation (in this case resulting from the half-difference) and $\alpha$ and $\beta$ fitting coefficients. The fitting coefficients are directly related to the level of cross-talk and therefore should be of the order of the level of cross-talk known for Narval.
\begin{figure}

	 \includegraphics[width=\columnwidth]{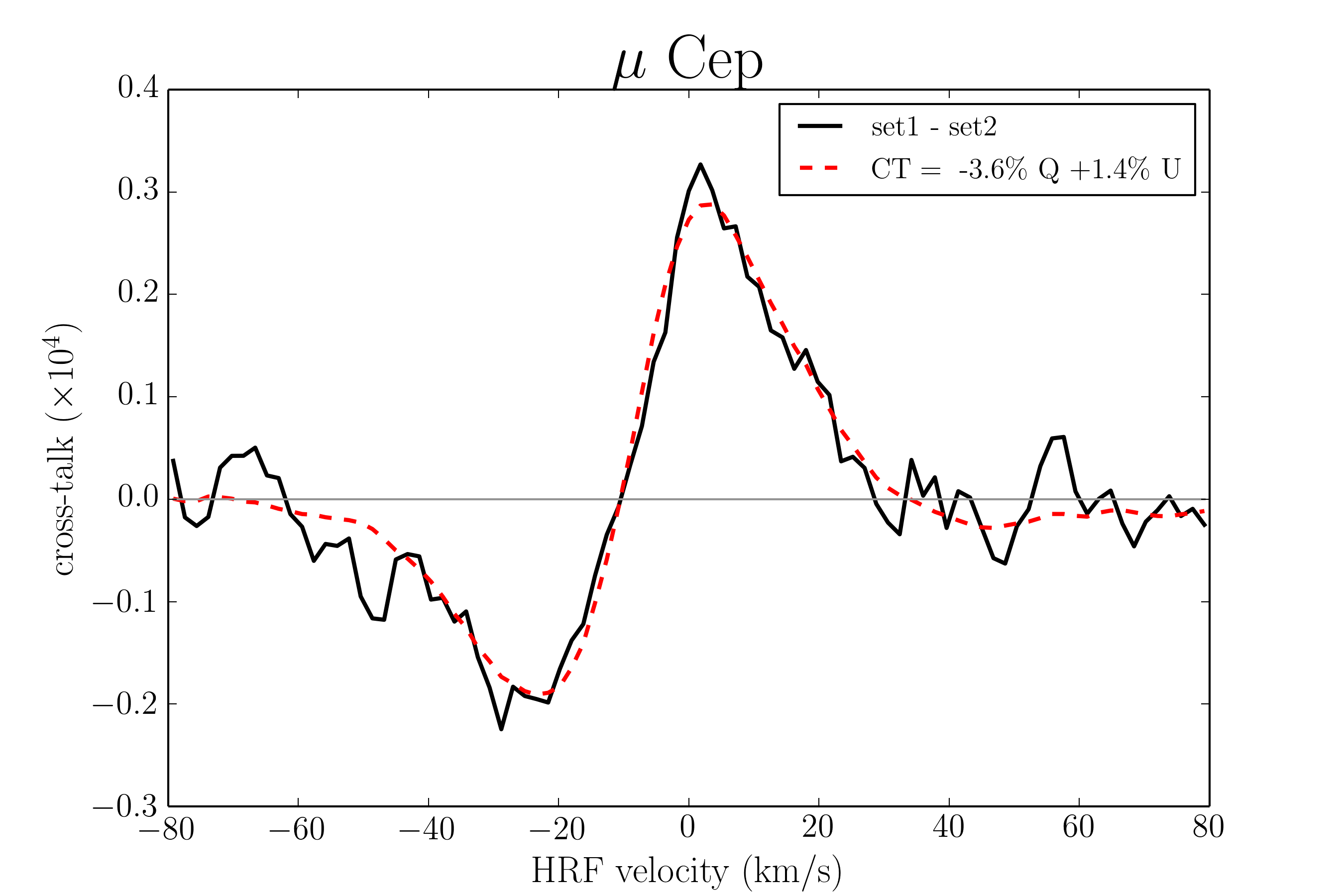}
 \caption{Least-squares fitting of $V_{\mathrm{cross-talk}}$ on the half-difference between  set 1 and set 2 (see Sect. \ref{CrossTalk}). The black solid line represents the Stokes~$V$ resulting from the half-difference and the dashed red line represents the least-squares fit of the  cross-talk function. The horizontal line represents 0 level.}
 \label{fig:figure 4}
\end{figure}

Figure \ref{fig:figure 4} shows the least-squares fitting of the spurious Stokes~$V$ with our cross-talk function. We notice that the fitted cross-talk function closely matches the signal.
In addition, the fitting coefficients $\alpha$ and $\beta$ are very consistent with the cross-talk level in Narval. 
The Stokes~Q and Stokes~U profiles of the January 2017 observation are identical to the ones resulting from the half-difference of December 2017 data. We therefore used the previous fitted cross-talk function to clean the observation of January 2017 (see Fig. \ref{fig:figure 5}).

\subsubsection{Possible origin of the null signal observed in $\mu$ Cep}
Regarding the null signal, we see in Figs. \ref{fig:figure 2} and \ref{fig:figure 3} that it changes its sign between set 1 and set 2, disappears in the sum, and is prominent in the difference, as is the case for Stokes Q and U. This behaviour demonstrates that the null signal on $\mu$ Cep, which was strong and ubiquitous in 2015 but weaker in 2016 and 2017 is also linked to the existence of the linear polarisation. Because its exact origin is not yet well established, in this work we only use our Stokes $V$ data free of null signal and corrected from cross-talk effect, namely our 2016 and 2017 observations, for magnetic study.
In the case of ${\rm \rho~Cas}$ and ${\rm \alpha^1~Her}$, where we detect low and no linear polarisation respectively, the cross-talk from linear polarisation does not contribute to the Stokes~$V$ signals down to the noise level.

\section{Results and discussion}\label{Results}

\subsection{Surface magnetic field estimates}\label{Bl}

We estimated the longitudinal component of the magnetic field, $<B_{\ell}>$, using the first order moment method \citep{1979A&A....74....1R} on the averaged LSD profiles. Although this method is adapted for dipolar fields with a classical Zeeman profile, meaning anti-symmetric Stokes~$V$ with respect to the line centre, in the case of non-classical Zeeman profiles it only gives a very rough estimate of the longitudinal magnetic field strength (see below).
Table \ref{tab3:table 3} summarises the values of $<B_{\ell}>$, the parameters $\overline{\lambda}$ and $\overline{g}$, and the statistical detection threshold in Stokes $V$ and in the null diagnosis for each target.

\begin{figure*}[h!]
        \begin{minipage}{\columnwidth}

	\includegraphics[width=\columnwidth]{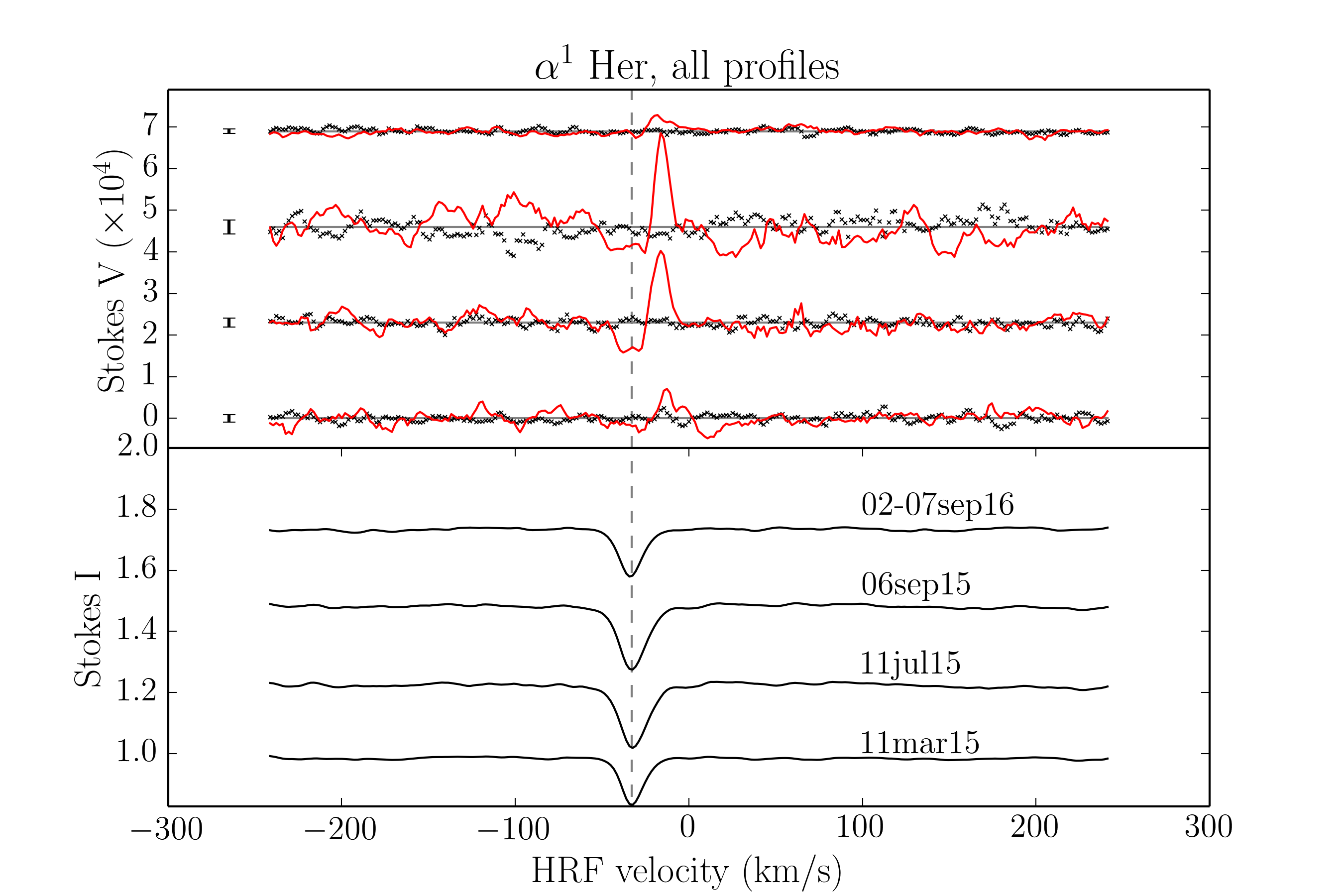}
	\includegraphics[width=\columnwidth]{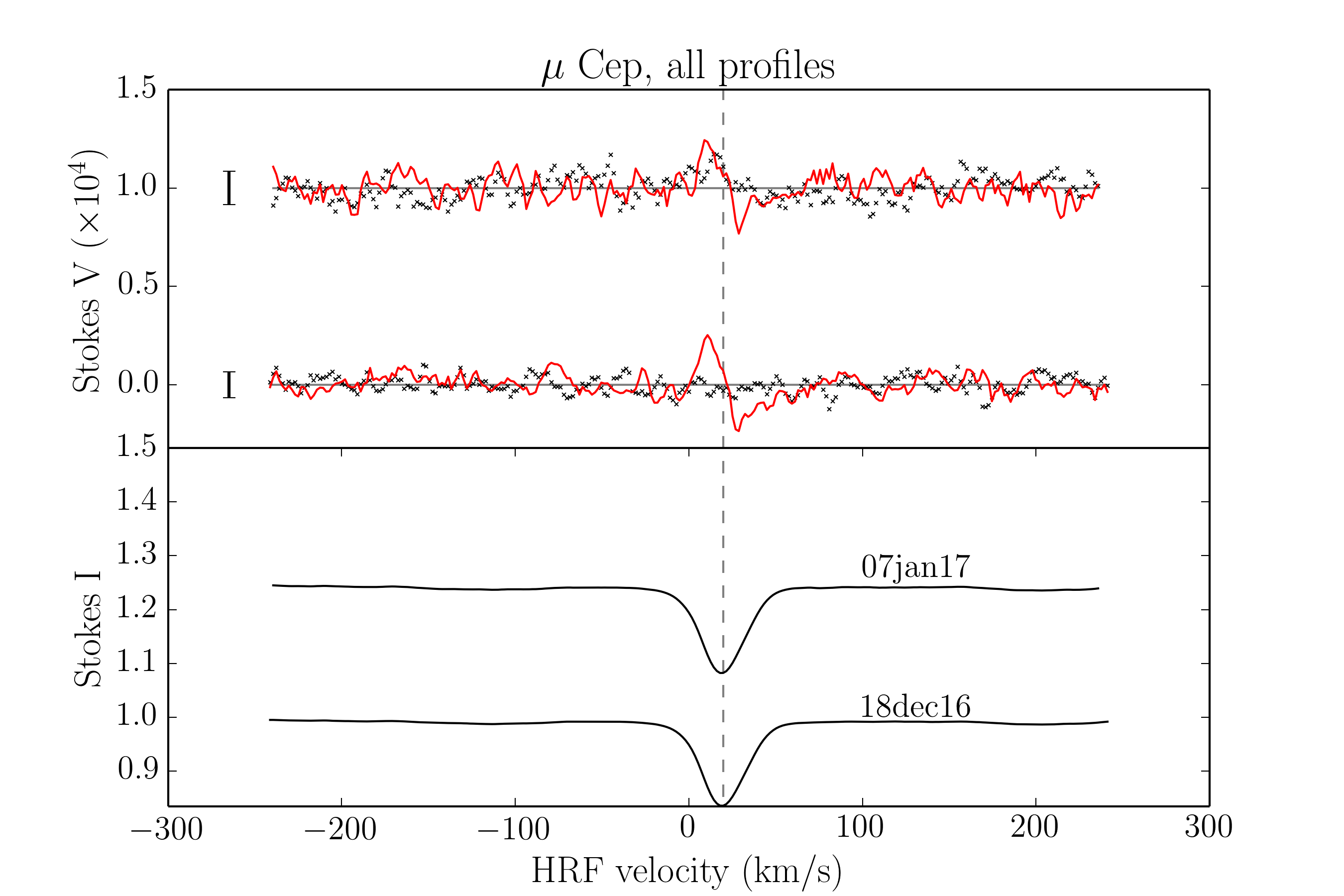}
        \end{minipage}
        \begin{minipage}{\columnwidth}
	
	\includegraphics[width=\columnwidth]{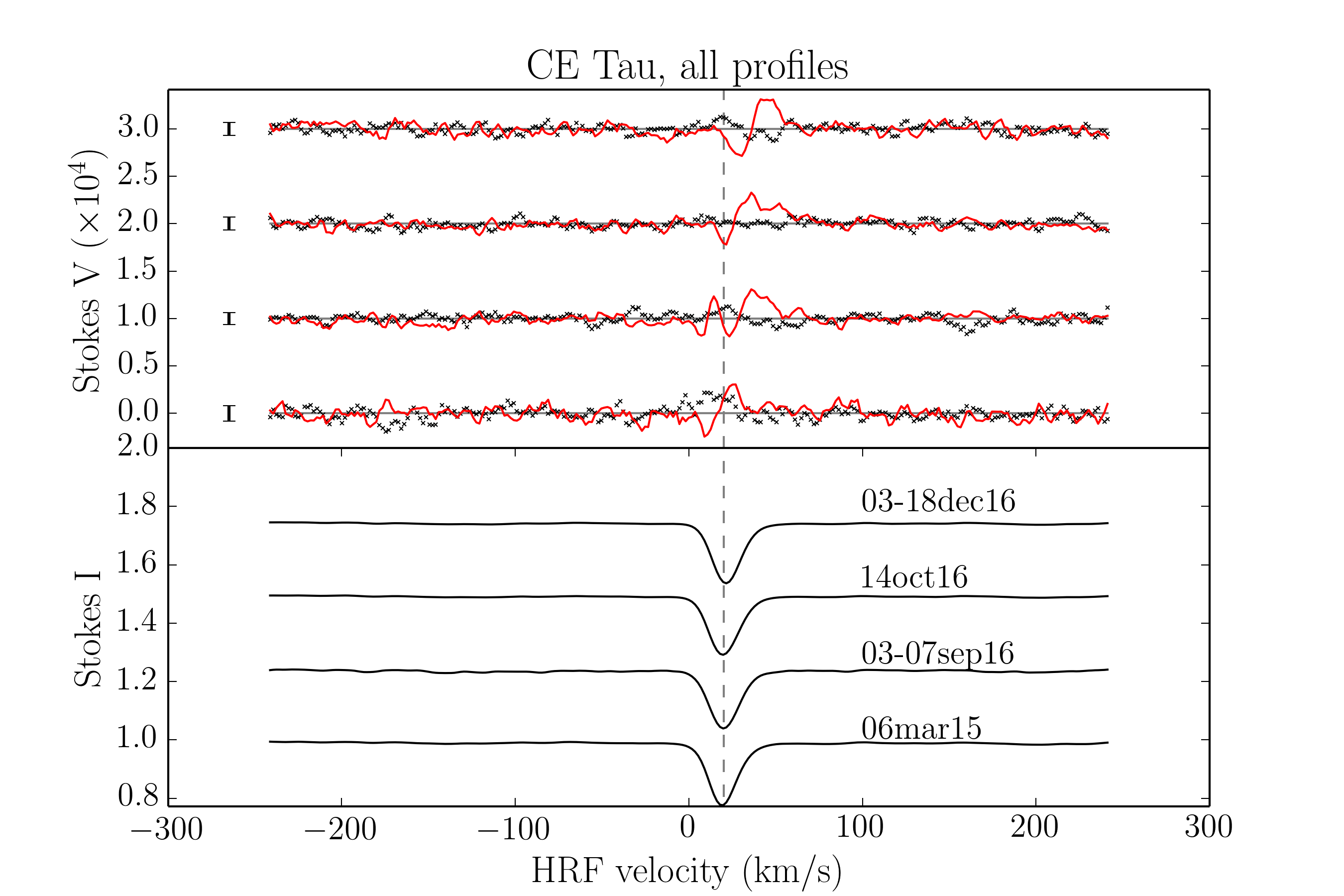}
	\includegraphics[width=\columnwidth]{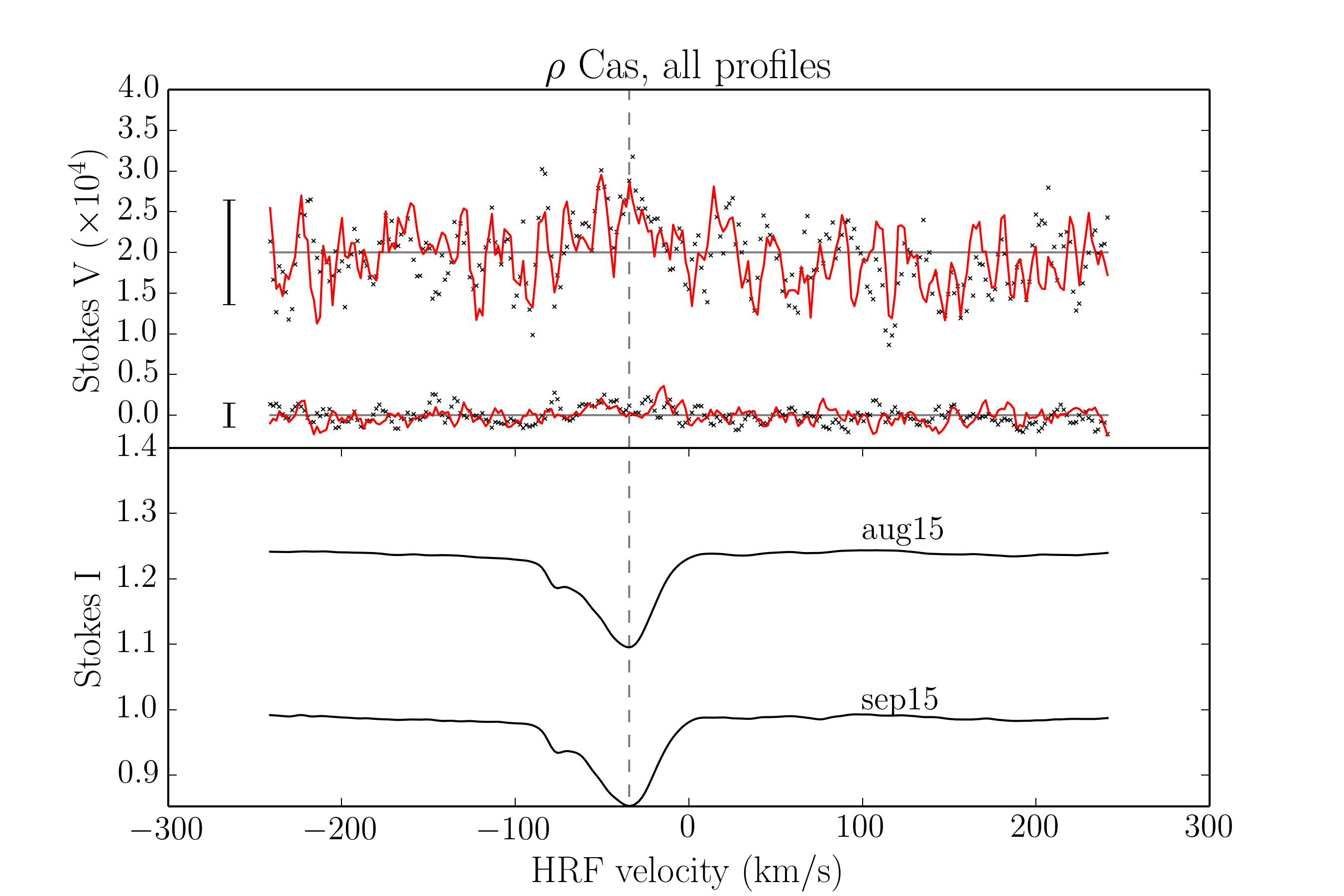}

	 \end{minipage}

    \caption{Variation of the LSD profiles of ${\rm \alpha^1~Her}$, ${\rm \mu~Cep}$, ${\rho~Cas}$ and CE Tau. For each observation point of Table \ref{tab2:table 2}, except for $\mu$~Cep, contiguous observations were averaged. For each star, the top panels represent all Stokes~$V$ profiles. The bottom panels represent the Stokes $I$ profiles. For each panel the dashed vertical lines mark the mean velocity of the star in the heliocentric rest frame, the horizontal solid lines the position of 0 and the black vertical segments the  mean 1$\sigma$ level. The null diagnosis is shown in each panel as black crosses.}
      \label{fig:figure 5}

\end{figure*}

 Figure \ref{fig:figure 5} presents the variations of the LSD profiles of our targets, but only for the cross-talk-free observations of $\mu$ Cep. Contiguous observations for each star were averaged. Very strong variations on time-scales consistent with atmospheric dynamics time-scales are observed for each target in Stokes~$V$ profiles. For Stokes~$I$ profiles, we notice a slight shift of the centre of the mean lines with time and a depth variation of the profiles, which are both consistent with photospheric motions \citep[see for instance ][]{2007A&A...469..671J}.

\cite{2010A&A...516L...2A} measure a weak magnetic field at the surface of ${\rm \alpha~Ori}$ at the Gauss-level. The amplitude of the Stokes $V$ for ${\rm \alpha~Ori}$ is at $ 0.3\times10^{-4}$ of the unpolarised continuum level, which is lower than the amplitude of the Stokes $V$ of ${\rm \alpha^1~Her}$, at maximum, $1.2\times10^{-4}$ of the continuum. 
However, the shape of the Stokes~$V$ we report for ${\rm \alpha^1~Her}$ (see Fig.~\ref{fig:figure 1}) is very similar to what is found on ${\rm \alpha~Ori}$ in \cite{2010A&A...516L...2A} with an asymmetric profile with two lobes of different depth. Indeed, in the case of ${\rm \alpha~Ori}$, one lobe of the Stokes $V$ profile is blue-shifted with respect to the centre of the Stokes $I$ profile, and we can see in Fig. \ref{fig:figure 1} that it is red-shifted in the case of ${\rm\alpha^1~Her}$. These Stokes~$V$ are very different from a classical Zeeman profile.
For CE Tau, the shapes of the Stokes $V$ profiles are very complex and have undergone strong variations since the beginning of the Large Program (Fig. \ref{fig:figure 5}). They are reminiscent of what is observed by \cite{2010MNRAS.408.2290G} on hotter supergiant stars. Besides, the same kind of non-classical Zeeman profiles have been reported by \cite{2016A&A...586A..97B} on Am stars.
There are several mechanisms that may be able to create such non-classical Zeeman profiles, for instance velocity fields. 
Cool evolved stars, AGB and RSG stars, undergo pulsations, shock waves and mass loss which in turn imply somewhat strong velocity fields at their surface and velocity gradients inside their atmosphere. It is sensible to consider that the shape of the Stokes $V$ profiles are partly due to these velocity fields.
Therefore we attribute the shape of the Stokes~$V$ profiles of our targets as mainly due to photospheric motions.

As explained in Sect. \ref{null} and Sect. \ref{CrossTalk}, most of our $\mu$~Cep observations are plagued by linear-to-circular polarisation cross-talk. For that reason we only considered as meaningful the two observations free of cross-talk. From these two observations a longitudinal magnetic field at and slightly below the Gauss-level is found.
For CE Tau and $\mu$~Cep the amplitudes of $ <B_{\ell}>$ are very similar to what is found in \cite{2010A&A...516L...2A} for $\alpha$~Ori.

As shown in Table \ref{tab3:table 3}, the estimates of the longitudinal component of the magnetic field of CE~Tau and ${\rm\alpha^1~Her}$  highlight a variation with time; $\mu$ Cep is not considered because of the cross-talk problem.
These variations occur on a typical time-scale of the order of several weeks for CE Tau and of the order of months for ${\rm\alpha^1~Her}$, although for the latter the observation sampling is very sparse compared to CE Tau.
These time-scales are fully compatible with what is found on ${\rm \alpha~Ori}$ \citep{2013EAS....60..161B} and on the magnetic AGB stars ${\rm \beta~Peg}$ and RZ Ari \citep{2013BlgAJ..19...14K}.
Indeed, using high quality spectropolarimetric observations, the presence of a surface magnetic field in a wide variety of cool evolved intermediate-mass stars has already been reported: \cite{2013BlgAJ..19...14K} have reported the detection of a significant surface field sometimes displaying time variations in M-type AGB stars . Also, \cite{2014A&A...561A..85L} have discovered a weak surface field, below the Gauss-level, in the pulsating variable Mira star ${\rm \chi~Cyg}$ (S spectral type).
Our results strengthen the idea that the presence of a surface magnetic field detectable at the Gauss-level or stronger is a widespread feature of cool evolved stars

\subsection{Magnetic field generation in cool evolved stars?}\label{MagGeneration}

The large-scale solar magnetic field is thought to be generated by a rotation driven dynamo \citep{2013SAAS...39.....C}.
A useful diagnostic to quantify influence of rotation on dynamo action is the Rossby number ($Ro$), which is the ratio between the rotational period and an averaged convective turnover time \citep{1984ApJ...279..763N,1984A&A...130..143M} that is used to assess the relative influence of inertia with respect to the Coriolis force. 
The most rapidly rotating cool stars exhibiting saturated coronal and chromospheric activity are characterised by an $Ro$ below about 0.1, whereas a somewhat slow rotator such as the Sun still exhibiting a large-scale rotation-dominated dynamo is characterised by $Ro\sim2$ \citep[see for instance][]{2011ApJ...743...48W}.
For ${\rm \alpha~Ori}$, the prototype of RSG stars, $Ro$ is about 90 \citep{2015IAUS..305..299J}. Because of this high an $Ro$, a large-scale dynamo generating a global magnetic field, such as the solar dynamo, is not expected to operate in RSG stars. However, simulations by \cite{2004A&A...423.1101D} and \cite{2003IAUS..210P.A12D} suggest that a small-scale dynamo, generating a magnetic field on the spatial scale of convective cells, could operate in RSG stars. Magneto-hydrodynamics (MHD) simulations by \cite{2004A&A...423.1101D} allow the existence of magnetic elements of strength up to 500 G, with small filling factors. This may result in detectable surface-averaged fields of few Gauss in good agreement with the detection of \cite{2010A&A...516L...2A} and with our present results. 
Moreover, by extrapolating magnetic field values from the CSE to the stellar surface, \cite{2005A&A...434.1029V} predict the same order of magnitude for the surface field of RSG stars. 

It is known that local turbulent fields exist in the Sun \citep[see][for a review]{2015SSRv..tmp...83S}.
Indeed the quiet photosphere is the site of dynamic magnetic activity, with a lifetime linked 
with that of  supergranules \citep{1997ApJ...487..424S}. These magnetic fields in quiet  regions are believed to 
be generated by local dynamo action driven by granular flows.
The large-scale convective motions modelled \citep{2002AN....323..213F} and reported \citep{2015EAS....71..243M} in the atmospheres 
of RSG stars may be compared to this solar supergranular pattern, which thus could generate a magnetic field because of turbulent motions.

\subsection{Individual cases}\label{Cases}
\subsubsection{${\rm \mu}$ Cep}\label{muCep}
The supergiant $\mu$ Cep (HD 206936) shares many similarities with ${\rm\alpha}$ Ori, in terms of spectroscopic properties (effective temperature / spectral type, 
surface gravity), although it may be more massive with  $M_\star \sim 25 M_{\mathrm{\odot}}$ (see Table \ref{tab1:table 1}).
According to the analysis performed 
by \cite{2007A&A...469..671J}, ${\rm \mu}$ Cep is convectively more active than ${\rm\alpha}$ Ori. It also loses mass at a slightly higher rate \citep[$\sim 2 \, 10^{-6}$ $M_{\mathrm{\odot}}$~yr$^{-1}$, with a possible decrease over the last $\sim 10^4$ yrs, see][and references therein]{2016AJ....151...51S}.

From our observations we have shown that the LSD statistics result in a "definite detection" both in Stokes $V$ and null profiles. 
As explained in Sect \ref{null} and \ref{CrossTalk} the "definite detection" of a signal in Stokes $V$, likely due to linear polarisation cross-talk, is very ambiguous and cannot be properly used.
The scattering polarisation that dominates the linearly polarised spectrum of ${\rm \mu~Cep}$, and also likely in ${\rm \alpha~Ori}$, makes the detection of weak magnetic fields very challenging. However, following the approach proposed by \cite{2009PASP..121..993B} we successfully measured a Stokes~$V$ signal of stellar origin and computed a longitudinal magnetic field of 1 Gauss. 
Besides, using set 1 and set 2 data (introduced in Sect. \ref{CrossTalk}) we cleaned, all spurious contributions from a new observation of $\mu$~Cep that was acquired shortly after set 1 and set 2. This new observation has also led to a non-ambiguous detection of a Zeeman signal and a longitudinal magnetic field strength at the Gauss-level.

However, circular polarisation measurement is not the only way of detecting magnetic fields.
Indeed the presence of a surface magnetic field could lead to some chromospheric activity.
We looked for hints of a chromospheric activity in Narval spectra and in IUE \citep[International Ultraviolet Explorer satellite, ][]{1978Natur.275..372B} spectra archives from SIMBAD. Firstly, in Narval spectra, the main indicators of chromospheric activity are the Ca II H\&K lines, $339.3~\mathrm{nm}~\mathrm{and}~339.5~\mathrm{nm}$, and the Ca II infrared triplet lines $849.8~\mathrm{nm},~854.2~\mathrm{nm}~\mathrm{and}~866.2~\mathrm{nm}$.
For active stars, emission can be seen in the core of these lines. \cite{2013LNP...857..231P} report no emission in the core of these lines for ${\rm \alpha~Ori}$ and no significant indication of activity. In our own study, we also found no sign of activity in these lines for $\mu$~Cep.

A large number of observations also exist for ${\rm \alpha~Ori}$ in IUE archives, from 1978 to 1992. We focused only on the data centred around the resonant lines of Mg II h\&k lines $\mathrm{h}~280~\mathrm{nm}~\mathrm{and}~\mathrm{k}~279.6~\mathrm{nm}$.
Strong emission is found in these resonant lines, suggesting chromospheric activity for ${\rm \alpha~Ori}$. Discussion on the properties and the detailed analysis of these lines can be found in the literature \citep[see for instance ][]{1978Natur.275..372B,1978MNRAS.183P..17B,1979ApJ...234.1023B,1980AcA....30..285G}. Ultraviolet observations of $\alpha$~Ori were performed by \cite{1998AJ....116.2501U} using the HST.
However, the number of such IUE observations is much lower for ${\rm \mu~Cep}$. Still, emission in the the Mg II resonant lines is found, but at a lower level that in ${\rm \alpha~Ori}$ \citep[see ][]{1986ApJ...308..859S}.
Even if the interpretation of this kind of data is beyond the scope of this paper, the emission found in the resonant lines of Mg II could be used as an indicator of chromospheric activity in ${\rm \mu~Cep}$.

\subsubsection{${\rm \alpha}$ Her}\label{alpHer}
The star ${\rm\alpha^1}$ Her (HD 156014) is the brightest component of a triple star system, the companions being the components of a close double-line 
spectroscopic binary, G8III + A9IV-V. Although ${\rm \alpha^1 ~Her}$ is classified as a M5Ib-II supergiant, its asteroseismic properties suggest that it is in fact a 2.5 $M_{\mathrm{\odot}}$ star \citep{2013AJ....146..148M}.
It may thus be instead an intermediate-mass  AGB star with an approximate age of 1.2 Gyr \citep{2011ASPC..445..163M,2013AJ....146..148M}.
Our measured magnetic field is also found to have a strength similar to the ones measured in AGB stars in \cite{2013BlgAJ..19...14K}, typically of several Gauss, meaning one order of magnitude higher than in ${\rm \alpha~Ori}$. 
The magnetic properties of ${\rm \alpha^1 ~Her}$ seem to be very similar of magnetic AGB stars and this is another argument favouring the star to be on the AGB, as suggested by \cite{2013AJ....146..148M}. 
Moreover, ${\rm \alpha^1 ~Her}$ is the only one from our RSG sample that does not show linear polarisation, as is also the case for non-pulsating AGB stars (L\`ebre et al. in preparation). This could be explained by the different surface and atmospheric dynamics between RSG and AGB stars.

The asymmetrical shape of the Stokes $V$, red-shifted with respect to the line centre of Stokes $I$ profile, is something that is known on ${\rm \alpha~Ori}$ and on other AGB stars, and it is likely linked to strong velocity gradients at the surface of the star, in the giant convective cells.
As in the case of ${\rm \mu~Cep}$, we searched for emission in the core of Ca II H\&K lines and in the Ca II infrared triplet lines tracing a chromospheric activity. 
However, we report no such emission in the core of these lines from our Narval spectra. Moreover no spectrum of ${\rm \alpha^1~Her}$ is available in the IUE archive.

\subsubsection{CE Tau}\label{ceTau}
The $\alpha$~Ori twin CE Tau (HD 36389) is a bright M2 Iab-Ib red supergiant star \citep{1998IBVS.4629....1W}.
We detected a surface magnetic field in CE Tau with a strength at about 1-2 Gauss, similar to what is known for ${\rm \alpha~Ori}$.

Moreover, the shape of the Stokes $V$ profiles, which strongly departs from a simple single-polarity Zeeman profile, is reminiscent of the complex signatures observed by \cite{2010MNRAS.408.2290G} on some hotter supergiant stars. The changes with time of the Stokes~$V$ profiles are again consistent with typical time-scale for surface dynamics  

Similar to ${\rm \alpha~Ori}$ and ${\rm \mu~Cep}$, CE Tau exhibits strong linear polarisation features, compared to circular polarisation, and this will be discussed in a future work. Contrary to the case of $\mu$~Cep, this linear polarisation is however not strong enough to result in any detectable cross-talk in the Stokes~$V$ LSD profiles of CE~Tau.  
Ultraviolet spectra from IUE archives show very intense emission in the resonant lines of Mg II, showing that CE Tau may be chromospherically active \citep[see for instance][]{1990ApJ...361..570H},  which is consistent with our observations. However, in our Narval spectra we do not detect any emission in the core of the Ca II H\&K lines and in the Ca II infrared triplet lines.

\subsubsection{${\rm \rho}$ Cas}\label{rhoCas}
The star $\rho$ Cas (HD 224014) is a massive F-type yellow supergiant, considered to be a post-RSG star as IRC +10420 is \citep{2016AJ....151...51S}.
 We do not detect any significant signatures in the Stokes~$V$ profiles of ${\rm \rho~Cas}$, noting that the feature seen in Fig. \ref{fig:figure 1} is below the "marginal detection" threshold, and the spectra on the  IUE archive show only little emission in the lines of Mg II. 
However, a non-detection of Stokes~$V$ signatures does not  necessarily mean an absence of surface magnetic field. The case of very weak magnetic field, below the Gauss-level,
or the case of complex fields that sometimes cancel out, can also lead to a non-detection.
Moreover, as has been reported on ${\rm \alpha~Ori}$, these fields can vary on a time scale of about a month, meaning much shorter than the very long rotational period. Therefore the surface magnetic field of  ${\rm \rho~Cas}$ may have simply been too weak or too complex to result in a detectable Stokes~$V$ signature at the two observation epochs. However, no emission in the core of the Ca II H\&K lines and in the Ca II infrared triplet lines is found.

\section{Summary and conclusions}\label{Conclusion}
With the spectropolarimeter Narval at TBL we observe four stars that are classified as late-type supergiants, ${\rm \mu~Cep}$, ${\rm \alpha^1~Her}$, CE Tau and ${\rm \rho~Cas}$, and we collect several high resolution spectra for each star.
Through the use of the LSD method we discover signatures in the LSD Stokes $V$ profiles of the three M-type supergiant stars of our sample, ${\rm \mu~Cep}$, ${\rm \alpha^1~Her}$, and CE Tau. Because RSG stars exhibit strong linear polarisation in spectral lines which is attributed to scattering processes \cite[see][]{2016A&A...591A.119A}, the possible contamination of the observed Stokes~$V$ LSD profiles by instrumental cross-talk must be carefully studied. In the case of $\mu$~Cep, the star with the strongest linear polarisation signatures in our sample, we show that the Stokes~$V$ LSD profile is heavily contaminated by cross-talk-induced spurious signals. We also attribute the nearly systematic presence of signals in the null diagnosis of $\mu$~Cep to cross-talk effects. However, we demonstrate that a clean cross-talk-free Stokes~$V$ LSD line profile can be recovered by combining two Stokes~$V$ polarimetric sequences collected with orthogonal instrument position angles. This approach, proposed by \cite{2009PASP..121..993B}, allows us to unambiguously detect and characterise the weak surface magnetic fields of RSG stars. We also establish that the level of linear polarisation in the two remaining RSG stars of our sample is not strong enough to result in any detectable cross-talk in our Stokes~$V$ LSD profiles. We can therefore safely attribute the detected, cleaned, Stokes~$V$ signals to the Zeeman effect and hence to the presence of surface magnetic fields.

We measure the weak surface longitudinal magnetic field of CE Tau and $\mu$~Cep and find amplitudes at the Gauss-level, fully consistent with what is known in ${\rm \alpha~Ori}$ the prototype of red supergiant stars. For the star ${\rm \alpha^1~Her}$, the measured magnetic field is greater than the field measured for the latter RSG stars, with a value consistent with what is known on cool AGB stars. This suggests the star belongs to AGB stars instead of the RSG stars, as already proposed by several authors.
We therefore show that $\alpha$~Ori is no more the only RSG star with a detected surface magnetic field.
Moreover, the variation time-scales of the surface field, detected both in CE Tau and ${\rm \alpha^1~Her}$, are in good agreement with small-scale dynamo
operating in giant convective cells lying at the surface of these cool evolved stars. Besides, emission in typical UV chromospheric lines has been found for these RSG stars consistent with a surface magnetism. 
We find no circularly polarised signature for the yellow post-supergiant star ${\rm \rho~Cas}$ at two different observation epochs. At that time, it is not clear whether we did not detect a surface magnetic field because the star has undergone a change in its photosphere, where the small-scale dynamo operates in RSG stars, during its evolution or because we observed it in a minimum of activity. More observations of post-RSG stars will help to understand the generation of surface magnetic fields at this evolutionary stage.
Observations of RSG stars with Narval are still running and confrontations with interferometric measurements are in progress. 
Besides, with the incoming infrared spectropolarimeters such as SPIP and SPIRou, we will have a better chance to measure, without ambiguities, the weak surface magnetic fields of these kind of stars. Indeed, although the Zeeman effect is stronger in the infrared with the amplitude of the Stokes~$V$ signatures being proportional to $\lambda$, the scattering linear polarisation, which scales with a ${\lambda^{-1}}$ law \citep{2016A&A...591A.119A}, is weaker in the infrared. For instance, the ratio between the level of continuum polarisation for a RSG star at 1500 nm and at 770 nm, to be presented in a forthcoming paper, is about 0.5, whereas this ratio for Zeeman polarisation is about 4.

\begin{acknowledgements}\label{Acknowledgements}
We thank the referee, Stefano Bagnulo, for his fruitful comments.
We acknowledge financial support from "Programme National de Physique Stellaire" (PNPS) of CNRS/INSU, France. We also acknowledge the TBL staff for providing service observing in Narval.
The spectropolarimetric data were reduced and analyzed using, respectively, the data reduction
software Libre-ESpRIT and the least-squares deconvolution routine (LSD),
written and provided by J.-F. Donati from IRAP-Toulouse (Observatoire Midi-
Pyrénées).
This research has made use of the SIMBAD database, operated at CDS (Strasbourg, France): \protect\url{http://simbad.u-strasbg.fr/}.
\end{acknowledgements}

\bibliographystyle{aa}
\bibliography{./tessore+17}

\begin{thebibliography}{56}
\expandafter\ifx\csname natexlab\endcsname\relax\def\natexlab#1{#1}\fi

\bibitem[{{Auri{\`e}re} {et~al.}(2010){Auri{\`e}re}, {Donati},
  {Konstantinova-Antova}, {Perrin}, {Petit}, \&
  {Roudier}}]{2010A&A...516L...2A}
{Auri{\`e}re}, M., {Donati}, J.-F., {Konstantinova-Antova}, R., {et~al.} 2010,
  \aap, 516, L2

\bibitem[{{Auri{\`e}re} {et~al.}(2015){Auri{\`e}re}, {Konstantinova-Antova},
  {Charbonnel}, {Wade}, {Tsvetkova}, {Petit}, {Dintrans}, {Drake}, {Decressin},
  {Lagarde}, {Donati}, {Roudier}, {Ligni{\`e}res}, {Schr{\"o}der},
  {Landstreet}, {L{\`e}bre}, {Weiss}, \& {Zahn}}]{2015A&A...574A..90A}
{Auri{\`e}re}, M., {Konstantinova-Antova}, R., {Charbonnel}, C., {et~al.} 2015,
  \aap, 574, A90

\bibitem[{{Auri{\`e}re} {et~al.}(2014){Auri{\`e}re}, {Konstantinova-Antova},
  {Espagnet}, {Petit}, {Roudier}, {Charbonnel}, {Donati}, \&
  {Wade}}]{2014IAUS..302..359A}
{Auri{\`e}re}, M., {Konstantinova-Antova}, R., {Espagnet}, O., {et~al.} 2014,
  in IAU Symposium, Vol. 302, Magnetic Fields throughout Stellar Evolution, ed.
  P.~{Petit}, M.~{Jardine}, \& H.~C. {Spruit}, 359--362

\bibitem[{{Auri{\`e}re} {et~al.}(2016){Auri{\`e}re}, {L{\'o}pez Ariste},
  {Mathias}, {L{\`e}bre}, {Josselin}, {Montarg{\`e}s}, {Petit}, {Chiavassa},
  {Paletou}, {Fabas}, {Konstantinova-Antova}, {Donati}, {Grunhut}, {Wade},
  {Herpin}, {Kervella}, {Perrin}, \& {Tessore}}]{2016A&A...591A.119A}
{Auri{\`e}re}, M., {L{\'o}pez Ariste}, A., {Mathias}, P., {et~al.} 2016, \aap,
  591, A119

\bibitem[{{Bagnulo} {et~al.}(2013){Bagnulo}, {Fossati}, {Kochukhov}, \&
  {Landstreet}}]{2013A&A...559A.103B}
{Bagnulo}, S., {Fossati}, L., {Kochukhov}, O., \& {Landstreet}, J.~D. 2013,
  \aap, 559, A103

\bibitem[{{Bagnulo} {et~al.}(2009){Bagnulo}, {Landolfi}, {Landstreet}, {Landi
  Degl'Innocenti}, {Fossati}, \& {Sterzik}}]{2009PASP..121..993B}
{Bagnulo}, S., {Landolfi}, M., {Landstreet}, J.~D., {et~al.} 2009, \pasp, 121,
  993

\bibitem[{{Barrick} {et~al.}(2010){Barrick}, {Benedict}, \&
  {Sabin}}]{2010SPIE.7735E..4CB}
{Barrick}, G., {Benedict}, T., \& {Sabin}, D. 2010, in \procspie, Vol. 7735,
  Ground-based and Airborne Instrumentation for Astronomy III, 77354C

\bibitem[{{Basri} \& {Linsky}(1979)}]{1979ApJ...234.1023B}
{Basri}, G.~S. \& {Linsky}, J.~L. 1979, \apj, 234, 1023

\bibitem[{{Bedecarrax} {et~al.}(2013){Bedecarrax}, {Petit}, {Auri{\`e}re},
  {Grunhut}, {Wade}, {Chiavassa}, {Donati}, {Konstantinova-Antova}, \&
  {Perrin}}]{2013EAS....60..161B}
{Bedecarrax}, I., {Petit}, P., {Auri{\`e}re}, M., {et~al.} 2013, in EAS
  Publications Series, Vol.~60, EAS Publications Series, ed. P.~{Kervella},
  T.~{Le Bertre}, \& G.~{Perrin}, 161--165

\bibitem[{{Bernat} \& {Lambert}(1978)}]{1978MNRAS.183P..17B}
{Bernat}, A.~P. \& {Lambert}, D.~L. 1978, \mnras, 183, 17P

\bibitem[{{Blaz{\`e}re} {et~al.}(2016){Blaz{\`e}re}, {Petit}, {Ligni{\`e}res},
  {Auri{\`e}re}, {Ballot}, {B{\"o}hm}, {Folsom}, {Gaurat}, {Jouve}, {Lopez
  Ariste}, {Neiner}, \& {Wade}}]{2016A&A...586A..97B}
{Blaz{\`e}re}, A., {Petit}, P., {Ligni{\`e}res}, F., {et~al.} 2016, \aap, 586,
  A97

\bibitem[{{Boggess} {et~al.}(1978){Boggess}, {Carr}, {Evans}, {Fischel},
  {Freeman}, {Fuechsel}, {Klinglesmith}, {Krueger}, {Longanecker}, \&
  {Moore}}]{1978Natur.275..372B}
{Boggess}, A., {Carr}, F.~A., {Evans}, D.~C., {et~al.} 1978, \nat, 275, 372

\bibitem[{{Boiarchuk} \& {Liubimkov}(1983)}]{1983IzKry..66..130B}
{Boiarchuk}, A.~A. \& {Liubimkov}, L.~S. 1983, Izvestiya Ordena Trudovogo
  Krasnogo Znameni Krymskoj Astrofizicheskoj Observatorii, 66, 130

\bibitem[{{Charbonneau}(2013)}]{2013SAAS...39.....C}
{Charbonneau}, P. 2013, Solar and Stellar Dynamos: Saas-Fee Advanced Course 39
  Swiss Society for Astrophysics and Astronomy, Saas-Fee Advanced Courses,
  Volume 39.~ISBN 978-3-642-32092-7.~Springer-Verlag Berlin Heidelberg, 2013,
  39

\bibitem[{{Charbonnel} {et~al.}(2017){Charbonnel}, {Decressin}, {Lagarde},
  {Gallet}, {Palacios}, {Auriere}, {Konstantinova-Antova}, {Mathis},
  {Anderson}, \& {Dintrans}}]{2017arXiv170310824C}
{Charbonnel}, C., {Decressin}, T., {Lagarde}, N., {et~al.} 2017, ArXiv e-prints
  [\eprint[arXiv]{1703.10824}]

\bibitem[{{Chiavassa} {et~al.}(2009){Chiavassa}, {Plez}, {Josselin}, \&
  {Freytag}}]{2009A&A...506.1351C}
{Chiavassa}, A., {Plez}, B., {Josselin}, E., \& {Freytag}, B. 2009, \aap, 506,
  1351

\bibitem[{{Danchi} {et~al.}(1994){Danchi}, {Bester}, {Degiacomi}, {Greenhill},
  \& {Townes}}]{1994AJ....107.1469D}
{Danchi}, W.~C., {Bester}, M., {Degiacomi}, C.~G., {Greenhill}, L.~J., \&
  {Townes}, C.~H. 1994, \aj, 107, 1469

\bibitem[{{Donati} {et~al.}(2006){Donati}, {Catala}, {Landstreet}, \&
  {Petit}}]{2006ASPC..358..362D}
{Donati}, J.-F., {Catala}, C., {Landstreet}, J.~D., \& {Petit}, P. 2006, in
  Astronomical Society of the Pacific Conference Series, Vol. 358, Astronomical
  Society of the Pacific Conference Series, ed. R.~{Casini} \& B.~W. {Lites},
  362

\bibitem[{{Donati} {et~al.}(1997){Donati}, {Semel}, {Carter}, {Rees}, \&
  {Collier Cameron}}]{1997MNRAS.291..658D}
{Donati}, J.-F., {Semel}, M., {Carter}, B.~D., {Rees}, D.~E., \& {Collier
  Cameron}, A. 1997, \mnras, 291, 658

\bibitem[{{Dorch}(2004)}]{2004A&A...423.1101D}
{Dorch}, S.~B.~F. 2004, \aap, 423, 1101

\bibitem[{{Dorch} \& {Freytag}(2003)}]{2003IAUS..210P.A12D}
{Dorch}, S.~B.~F. \& {Freytag}, B. 2003, in IAU Symposium, Vol. 210, Modelling
  of Stellar Atmospheres, ed. N.~{Piskunov}, W.~W. {Weiss}, \& D.~F. {Gray},
  A12

\bibitem[{{Folsom} {et~al.}(2016){Folsom}, {Petit}, {Bouvier}, {L{\`e}bre},
  {Amard}, {Palacios}, {Morin}, {Donati}, {Jeffers}, {Marsden}, \&
  {Vidotto}}]{2016MNRAS.457..580F}
{Folsom}, C.~P., {Petit}, P., {Bouvier}, J., {et~al.} 2016, \mnras, 457, 580

\bibitem[{{Freytag} {et~al.}(2002){Freytag}, {Steffen}, \&
  {Dorch}}]{2002AN....323..213F}
{Freytag}, B., {Steffen}, M., \& {Dorch}, B. 2002, Astronomische Nachrichten,
  323, 213

\bibitem[{{Glebocki} \& {Stawikowski}(1980)}]{1980AcA....30..285G}
{Glebocki}, R. \& {Stawikowski}, A. 1980, \actaa, 30, 285

\bibitem[{{Grunhut} {et~al.}(2010){Grunhut}, {Wade}, {Hanes}, \&
  {Alecian}}]{2010MNRAS.408.2290G}
{Grunhut}, J.~H., {Wade}, G.~A., {Hanes}, D.~A., \& {Alecian}, E. 2010, \mnras,
  408, 2290

\bibitem[{{Haisch} {et~al.}(1990){Haisch}, {Bookbinder}, {Maggio}, {Vaiana}, \&
  {Bennett}}]{1990ApJ...361..570H}
{Haisch}, B.~M., {Bookbinder}, J.~A., {Maggio}, A., {Vaiana}, G.~S., \&
  {Bennett}, J.~O. 1990, \apj, 361, 570

\bibitem[{{Haubois} {et~al.}(2009){Haubois}, {Perrin}, {Lacour}, {Verhoelst},
  {Meimon}, {Mugnier}, {Thi{\'e}baut}, {Berger}, {Ridgway}, {Monnier},
  {Millan-Gabet}, \& {Traub}}]{2009A&A...508..923H}
{Haubois}, X., {Perrin}, G., {Lacour}, S., {et~al.} 2009, \aap, 508, 923

\bibitem[{{H{\"o}fner}(2008)}]{2008A&A...491L...1H}
{H{\"o}fner}, S. 2008, \aap, 491, L1

\bibitem[{{Josselin} {et~al.}(2015){Josselin}, {Lambert}, {Auri{\`e}re},
  {Petit}, \& {Ryde}}]{2015IAUS..305..299J}
{Josselin}, E., {Lambert}, J., {Auri{\`e}re}, M., {Petit}, P., \& {Ryde}, N.
  2015, in IAU Symposium, Vol. 305, Polarimetry, ed. K.~N. {Nagendra},
  S.~{Bagnulo}, R.~{Centeno}, \& M.~{Jes{\'u}s Mart{\'{\i}}nez Gonz{\'a}lez},
  299--300

\bibitem[{{Josselin} \& {Plez}(2007)}]{2007A&A...469..671J}
{Josselin}, E. \& {Plez}, B. 2007, \aap, 469, 671

\bibitem[{{Konstantinova-Antova} {et~al.}(2013){Konstantinova-Antova},
  {Aur{\'{\i}}ere}, {Charbonnel}, {Wade}, {Kolev}, {Antov}, {Tsvetkova},
  {Schr{\"o}der}, {Drake}, {Petit}, {de Medeiros}, {L{\'e}bre}, {Zhilyaev},
  {Verlyuk}, {Svyatogorov}, {Gershberg}, {Lovkaya}, {Bogdanovski}, {Stateva},
  {Cabanac}, {Avgoloupis}, {Contadakis}, \& {Seiradakis}}]{2013BlgAJ..19...14K}
{Konstantinova-Antova}, R., {Aur{\'{\i}}ere}, M., {Charbonnel}, C., {et~al.}
  2013, Bulgarian Astronomical Journal, 19, 14

\bibitem[{{Kupka} {et~al.}(1999){Kupka}, {Piskunov}, {Ryabchikova}, {Stempels},
  \& {Weiss}}]{1999A&AS..138..119K}
{Kupka}, F., {Piskunov}, N., {Ryabchikova}, T.~A., {Stempels}, H.~C., \&
  {Weiss}, W.~W. 1999, \aaps, 138, 119

\bibitem[{{Kurucz}(2005)}]{2005MSAIS...8...14K}
{Kurucz}, R.~L. 2005, Memorie della Societa Astronomica Italiana Supplementi,
  8, 14

\bibitem[{{L{\`e}bre} {et~al.}(2014){L{\`e}bre}, {Auri{\`e}re}, {Fabas},
  {Gillet}, {Herpin}, {Konstantinova-Antova}, \& {Petit}}]{2014A&A...561A..85L}
{L{\`e}bre}, A., {Auri{\`e}re}, M., {Fabas}, N., {et~al.} 2014, \aap, 561, A85

\bibitem[{{Levesque} {et~al.}(2005){Levesque}, {Massey}, {Olsen}, {Plez},
  {Josselin}, {Maeder}, \& {Meynet}}]{2005ApJ...628..973L}
{Levesque}, E.~M., {Massey}, P., {Olsen}, K.~A.~G., {et~al.} 2005, \apj, 628,
  973

\bibitem[{{Mangeney} \& {Praderie}(1984)}]{1984A&A...130..143M}
{Mangeney}, A. \& {Praderie}, F. 1984, \aap, 130, 143

\bibitem[{{Montarg{\`e}s} {et~al.}(2015){Montarg{\`e}s}, {Kervella}, {Perrin},
  {Chiavassa}, \& {Auri{\`e}re}}]{2015EAS....71..243M}
{Montarg{\`e}s}, M., {Kervella}, P., {Perrin}, G., {Chiavassa}, A., \&
  {Auri{\`e}re}, M. 2015, in EAS Publications Series, Vol.~71, EAS Publications
  Series, 243--247

\bibitem[{{Moravveji} {et~al.}(2013){Moravveji}, {Guinan}, {Khosroshahi}, \&
  {Wasatonic}}]{2013AJ....146..148M}
{Moravveji}, E., {Guinan}, E.~F., {Khosroshahi}, H., \& {Wasatonic}, R. 2013,
  \aj, 146, 148

\bibitem[{{Moravveji} {et~al.}(2011){Moravveji}, {Guinan}, \&
  {Sobouti}}]{2011ASPC..445..163M}
{Moravveji}, E., {Guinan}, E.~F., \& {Sobouti}, Y. 2011, in Astronomical
  Society of the Pacific Conference Series, Vol. 445, Why Galaxies Care about
  AGB Stars II: Shining Examples and Common Inhabitants, ed. F.~{Kerschbaum},
  T.~{Lebzelter}, \& R.~F. {Wing}, 163

\bibitem[{{Morin} {et~al.}(2008){Morin}, {Donati}, {Forveille}, {Delfosse},
  {Dobler}, {Petit}, {Jardine}, {Collier Cameron}, {Albert}, {Manset},
  {Dintrans}, {Chabrier}, \& {Valenti}}]{2008MNRAS.384...77M}
{Morin}, J., {Donati}, J.-F., {Forveille}, T., {et~al.} 2008, \mnras, 384, 77

\bibitem[{{Noyes} {et~al.}(1984){Noyes}, {Hartmann}, {Baliunas}, {Duncan}, \&
  {Vaughan}}]{1984ApJ...279..763N}
{Noyes}, R.~W., {Hartmann}, L.~W., {Baliunas}, S.~L., {Duncan}, D.~K., \&
  {Vaughan}, A.~H. 1984, \apj, 279, 763

\bibitem[{{Petit} {et~al.}(2013){Petit}, {Auri{\`e}re}, {Konstantinova-Antova},
  {Morgenthaler}, {Perrin}, {Roudier}, \& {Donati}}]{2013LNP...857..231P}
{Petit}, P., {Auri{\`e}re}, M., {Konstantinova-Antova}, R., {et~al.} 2013, in
  Lecture Notes in Physics, Berlin Springer Verlag, Vol. 857, Lecture Notes in
  Physics, Berlin Springer Verlag, ed. J.-P. {Rozelot} \& C.~. {Neiner}, 231

\bibitem[{{Petit} {et~al.}(2010){Petit}, {Ligni{\`e}res}, {Wade},
  {Auri{\`e}re}, {B{\"o}hm}, {Bagnulo}, {Dintrans}, {Fumel}, {Grunhut},
  {Lanoux}, {Morgenthaler}, \& {Van Grootel}}]{2010A&A...523A..41P}
{Petit}, P., {Ligni{\`e}res}, F., {Wade}, G.~A., {et~al.} 2010, \aap, 523, A41

\bibitem[{{Rees} \& {Semel}(1979)}]{1979A&A....74....1R}
{Rees}, D.~E. \& {Semel}, M.~D. 1979, \aap, 74, 1

\bibitem[{{Sabin} {et~al.}(2014){Sabin}, {Wade}, \&
  {L{\`e}bre}}]{2014arXiv1410.6224S}
{Sabin}, L., {Wade}, G.~A., \& {L{\`e}bre}, A. 2014, ArXiv e-prints
  [\eprint[arXiv]{1410.6224}]

\bibitem[{{Schrijver} {et~al.}(1997){Schrijver}, {Title}, {van Ballegooijen},
  {Hagenaar}, \& {Shine}}]{1997ApJ...487..424S}
{Schrijver}, C.~J., {Title}, A.~M., {van Ballegooijen}, A.~A., {Hagenaar},
  H.~J., \& {Shine}, R.~A. 1997, \apj, 487, 424

\bibitem[{{Schwarzschild}(1975)}]{1975ApJ...195..137S}
{Schwarzschild}, M. 1975, \apj, 195, 137

\bibitem[{{Shenoy} {et~al.}(2016){Shenoy}, {Humphreys}, {Jones}, {Marengo},
  {Gehrz}, {Helton}, {Hoffmann}, {Skemer}, \& {Hinz}}]{2016AJ....151...51S}
{Shenoy}, D., {Humphreys}, R.~M., {Jones}, T.~J., {et~al.} 2016, \aj, 151, 51

\bibitem[{{Silvester} {et~al.}(2012){Silvester}, {Wade}, {Kochukhov},
  {Bagnulo}, {Folsom}, \& {Hanes}}]{2012MNRAS.426.1003S}
{Silvester}, J., {Wade}, G.~A., {Kochukhov}, O., {et~al.} 2012, \mnras, 426,
  1003

\bibitem[{{Stencel} {et~al.}(1986){Stencel}, {Carpenter}, \&
  {Hagen}}]{1986ApJ...308..859S}
{Stencel}, R.~E., {Carpenter}, K.~G., \& {Hagen}, W. 1986, \apj, 308, 859

\bibitem[{{Stenflo}(2015)}]{2015SSRv..tmp...83S}
{Stenflo}, J.~O. 2015, \ssr [\eprint[arXiv]{1508.03312}]

\bibitem[{{Uitenbroek} {et~al.}(1998){Uitenbroek}, {Dupree}, \&
  {Gilliland}}]{1998AJ....116.2501U}
{Uitenbroek}, H., {Dupree}, A.~K., \& {Gilliland}, R.~L. 1998, \aj, 116, 2501

\bibitem[{{Vlemmings}(2014)}]{2014IAUS..302..389V}
{Vlemmings}, W.~H.~T. 2014, in IAU Symposium, Vol. 302, Magnetic Fields
  throughout Stellar Evolution, ed. P.~{Petit}, M.~{Jardine}, \& H.~C.
  {Spruit}, 389--397

\bibitem[{{Vlemmings} {et~al.}(2005){Vlemmings}, {van Langevelde}, \&
  {Diamond}}]{2005A&A...434.1029V}
{Vlemmings}, W.~H.~T., {van Langevelde}, H.~J., \& {Diamond}, P.~J. 2005, \aap,
  434, 1029

\bibitem[{{Wasatonic} \& {Guinan}(1998)}]{1998IBVS.4629....1W}
{Wasatonic}, R. \& {Guinan}, E.~F. 1998, Information Bulletin on Variable
  Stars, 4629

\bibitem[{{Wright} {et~al.}(2011){Wright}, {Drake}, {Mamajek}, \&
  {Henry}}]{2011ApJ...743...48W}
{Wright}, N.~J., {Drake}, J.~J., {Mamajek}, E.~E., \& {Henry}, G.~W. 2011,
  \apj, 743, 48

\end{thebibliography}

\end{document}